\documentclass[preprint2]{emulateapj}

\usepackage{hyperref}
\usepackage{color}

\def\hii{H {\sc ii}~}

\begin{document}

\title{Optical photometric and polarimetric investigation of NGC\,1931}

\author{A. K. PANDEY\altaffilmark{1}, C. ESWARAIAH\altaffilmark{1}, SAURABH SHARMA\altaffilmark{1}, M. R. SAMAL\altaffilmark{2}, N. CHAUHAN\altaffilmark{3}, W. P. CHEN\altaffilmark{3}, J. JOSE\altaffilmark{4}, D. K. OJHA\altaffilmark{5}, RAM KESH YADAV\altaffilmark{1} AND H. C. CHANDOLA\altaffilmark{6}}
\altaffiltext{1}{Aryabhatta Research Institute of Observational Sciences, Manora Peak, Nainital 263129, India}
\altaffiltext{2}{Laboratoire d'Astrophysique de Marseille~-~LAM, Universit\'{e} d'Aix-Marseille $\&$ CNRS, UMR7326, 13388 Marseille Cedex 13, France}
\altaffiltext{3}{Institute of Astronomy, National Central University, Chung-Li 32054, Taiwan}
\altaffiltext{4}{Indian Institute of Astrophysics, Koramangala, Banglore 560034, India}
\altaffiltext{5}{Tata Institute of Fundamental Research, Mumbai 400 005, India}
\altaffiltext{6}{Department of Physics, Kumaun University, Nainital 263129, India}
\email{pandey@aries.res.in}


\begin{abstract}

We present optical photometric and polarimetric observations of stars towards NGC\,1931 with the aim to 
derive the cluster parameters such as distance, reddening, age and luminosity/mass function as well as to 
understand the dust properties and star formation in the region. The distance to the cluster is found to be 
2.3$\pm$0.3 kpc and the reddening $E(B-V)$ in the region is found to be variable. The stellar density contours 
reveal two clustering in the region. The observations suggest differing reddening law within the cluster region. 
Polarization efficiency of the dust grains towards the direction of the cluster is found to be less than that for 
the general diffuse interstellar medium (ISM). The slope of the mass function (-0.98$\pm$0.22) in the southern region 
in the mass range 0.8~\textless~$M/M_{\sun}$~$\textless$~9.8 is found to be shallower in comparison to that in 
the northern region (-1.26$\pm$0.23), which is comparable to the Salpeter value (-1.35). The K-band luminosity 
function (KLF) of the region is found to be comparable to the average value of slope ($\sim$0.4) for young clusters 
obtained by \citet*{2003ARA&A..41...57L}, however, the slope of the KLF is steeper in the northern region as compared 
to the southern region. The region is probably ionized by two B2 main-sequence type stars. The mean age of the young 
stellar objects (YSOs) is found to be 2$\pm$1 Myr which suggests that the identified  YSOs could be younger than the 
ionizing sources of the region. The morphology of the region, the distribution of the YSOs as well as ages of the YSOs 
and ionizing sources indicate a triggered star formation in the region. 

\end{abstract}


\section{INTRODUCTION}

It is believed that the dust grains in the interstellar medium (ISM) and intra-cluster medium(ICM) 
are aligned due to the local magnetic field. The light passing through these mediums gets 
linearly polarized at a level of few percent. Thus the polarimetry is an efficient tool to 
study the properties of the dust grains, magnetic field orientation, nature of extinction law etc along a  line of sight.  
The polarimetric observations towards  young star clusters which are still embedded in the parent molecular cloud are 
of special interest as many basic parameters like membership, distance, age, color excess $[E(B-V)]$ etc. for theses regions are 
known with relatively better accuracy, which helps in analyzing the polarimetric data with a better confidence. 
The ultraviolet (UV) radiation due to the massive members in  these regions have strong impact on the ICM and 
the dust grains in the ICM can undergo destruction processes due to the radiation pressure, grain-grain collisions, 
sputtering or shattering, etc. Consequently the dust grain size in the ICM could be smaller than the mean value for 
the diffuse ISM. The study of interaction of the ICM dust grains with the local magnetic field may provide crucial 
clues to understand the physical processes (e.g., role of magnetic field in the initial cloud collapse) 
acting in such environments.

The study of star formation process and stellar evolution is another basic problem in astrophysics. 
Since it is believed that the majority of the stars in our Galaxy are formed in groups known as star clusters, 
the  star clusters are useful objects to study the star formation process. The initial distribution of stellar masses 
i.e, the initial mass function ($IMF$) is one of the basic tools to understand the formation and evolution of 
stellar systems. Since the young clusters (age $\textless$~10 Myr) are assumed to be less affected by the dynamical 
evolution, their mass function ($MF$) can be considered as the $IMF$. Thus the young clusters also serve as 
ideal laboratories  to study  the form of $IMF$ and its variation within space and time. 

The paper is continuation of our efforts to understand the star formation scenario 
\citep{2008MNRAS.383.1241P,2007MNRAS.380.1141S,2012arXiv1204.2897S,2008MNRAS.384.1675J,2011MNRAS.411.2530J,2007ApJ...671..555S,2010ApJ...714.1015S,2009MNRAS.396..964C,2011PASJ...63..795C,2011MNRAS.415.1202C}  
and dust characteristics in star-forming regions as well as to study the structure of the magnetic field in various environments of the Galaxy 
\citep[][hereafter E11 and E12, respectively]{2011MNRAS.411.1418E,2012MNRAS.419.2587E}. In this paper we report results based on broad-band optical 
photometric and polarimetric observations around the cluster NGC\,1931. 
We have also used archived near infrared ($NIR$) and mid infrared ($MIR$) data.

NGC\,1931 ($\alpha_{2000}$ = 05$^{h}$ 31$^{m}$ 25$^{s}$, 
$\delta_{2000}$ = +34$\degr$~14$\arcmin$~42$\arcsec$;  $l$=$173\fdg9$, $b$=$0\fdg28$) 
is a young star cluster associated with a gas-dust complex and the bright 
nebula Sh2-237 in Auriga. The distance estimates for the cluster vary between 1.8 kpc and 3.1 kpc and the 
post-main-sequence age of the cluster is reported to be $\sim$ 10 Myr 
\citep{1979A&AS...38..197M,1986Ap&SS.120..107P,1994BASI...22..291B,2004AJ....128.2306C,2009MNRAS.397.1915B}.
\citet{1975A&A....39..481G} 
reported that Sh2-237 is excited by a star of spectral type B0.5. \citet{1989ApJS...70..731L} 
have found that a portion of Sh2-237 is obscured by a molecular cloud.

Fig. 1 shows color-composite images of the Sh2-237 region, obtained using 2 micron all sky-survey (2MASS) 
$K_{s}$ (blue), {\it Spitzer IRAC} 
3.6 $\mu$m (green) and 4.5 $\mu$m (red) images (left panel), and ${\rm H\alpha}$ (green) and 3.6 $\mu$m (red) images 
(right panel), which suggest that the Sh2-237 is a bright optical \hii region of diameter $\sim$ 4 arcmin and  
is surrounded by a dust ring, revealed by the MIR emission in the {\it Spitzer IRAC} channels 
1 and 2 (centered at 3.6 $\mu$m and 4.5 $\mu$m). 

In Sections \ref{data_reduction} and \ref{archival_data}, we describe the observations, data reduction and 
archived data used in the present study. In  Sections  \ref{structure}$-$\ref{IMF_KLF}, we describe the results obtained. 
The star formation scenario in the NGC\,1931 (Sh2-237) region is described in Section \ref{SFR_SCENARIO}. 
The results are summarized in Section \ref{conclusions}.
 
\section{OBSERVATIONS AND  DATA REDUCTION} \label{data_reduction}

\subsection{Optical photometric data}

\begin{figure*}
\centering
\resizebox{8.175cm}{8.175cm}{\includegraphics{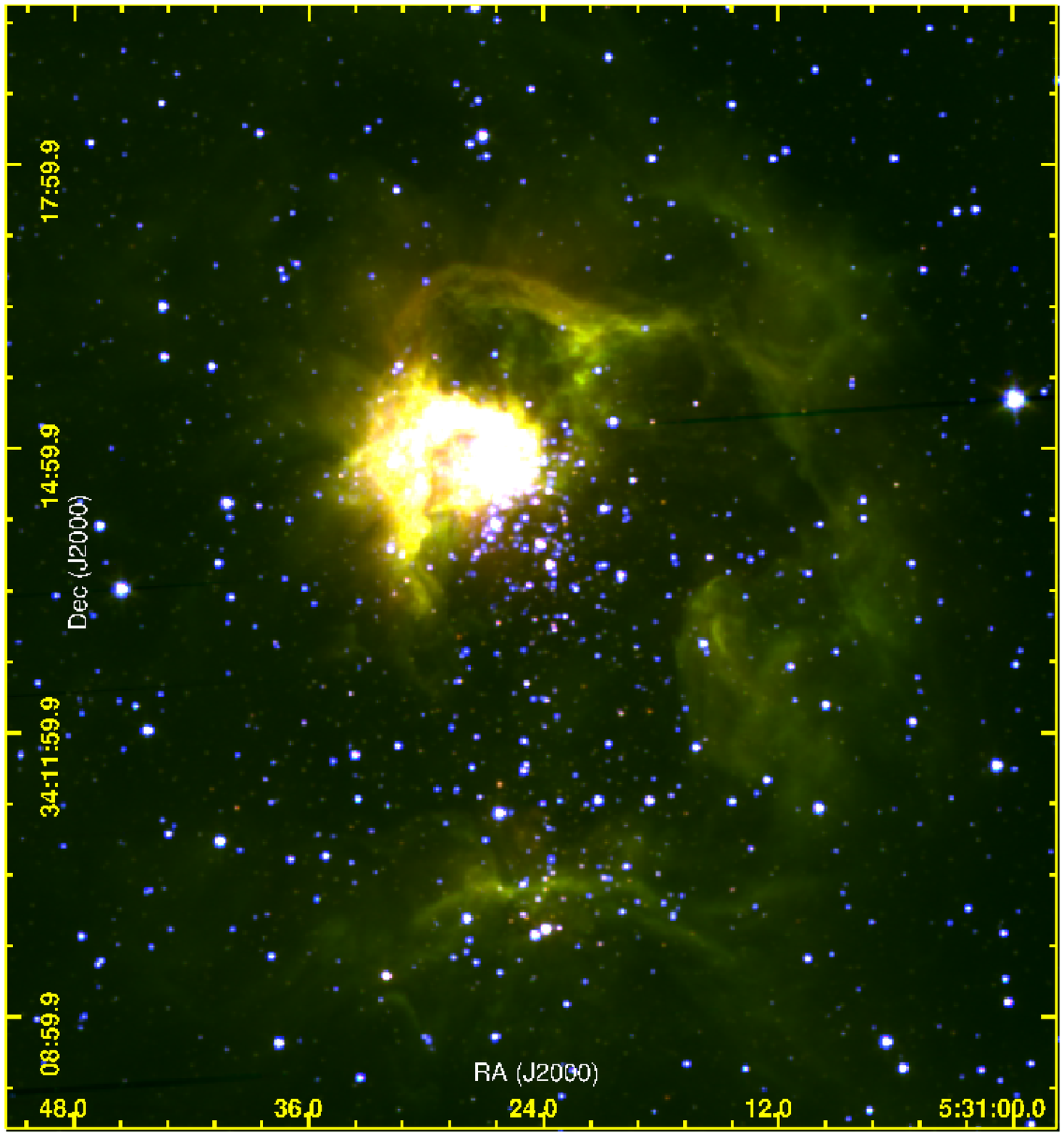}}
\resizebox{8.175cm}{8.175cm}{\includegraphics{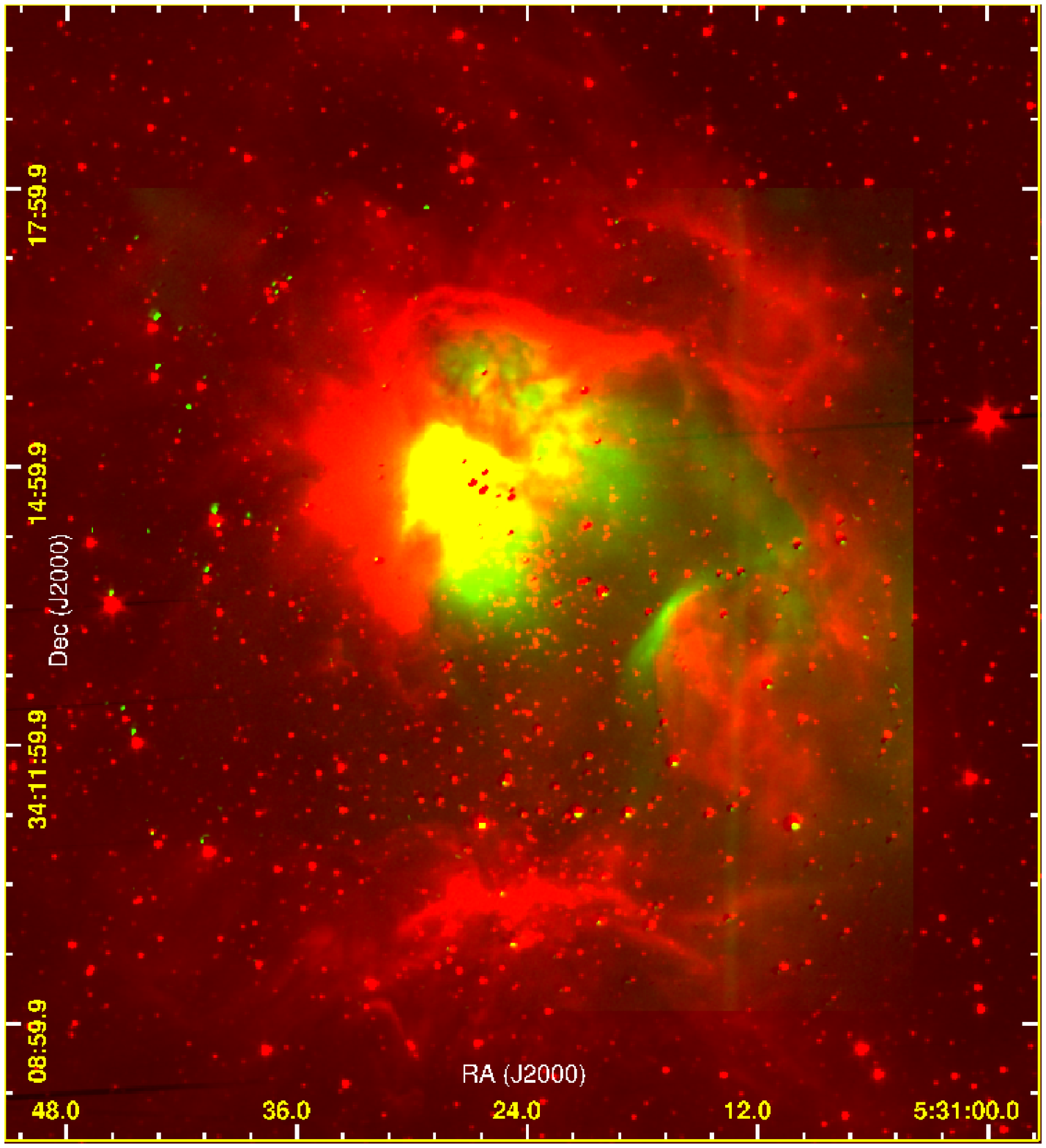}}
\caption{{\it (left panel:)} The color-composite images obtained using the Ks (blue), 3.6 $\mu$m (green) 
and 4.5 $\mu$m (red) for an area of $\simeq $15 $\times$ 15 arcmin$^2$ around NGC\,1931. 
{\it (right panel:)} Same as the left panel but using 
the $H\alpha$ (green) and 3.6 $\mu$m (red) (see the electronic version for the color image).}
\label{color_composite}
\end{figure*}

\begin{table}
\centering
\scriptsize
\caption{\label{log} Log of observations.}
\begin{tabular}{@{}rrr@{}}
\hline
Date of observation & Filter& Exp. (sec)$\times$ No. of frames\\
\hline
2005 December 31 & $B$   &  $1200\times3,120\times1$\\
	& $V$   &  $900\times3,60\times1$\\
	& $I_{c}$   &  $300\times2$\\
	& $R_{c}$   &  $300\times4$\\
	& $U$   &  $1800\times4,300\times1$\\
2006 January 22 & $I_{c}$   &  $300\times4,60\times3$\\
2006 January 23	& $U$   &  $300\times3$\\
	& $B$   &  $180\times3,60\times3$\\
	& $V$   &  $120\times3,30\times3$\\
	& $I_{c}$   &  $60\times4,20\times3$\\
	& $R_{c}$   &  $60\times3,20\times3$\\
2006 February 24 & $H{\alpha}$  &  $900\times2,300\times1$\\
	& $continuum$   &  $900\times2,300\times1$\\
\hline
\end{tabular}
\end{table}

The photometric data were acquired on 2005 December 31; 2006 January 22, 23; and 2006 February 24, using the 2048 $\times$ 2048 
pixel CCD camera mounted on the f/13 Cassegrain focus of the 104-cm Sampurnanand Telescope of Aryabhatta Research 
Institute of Observational Sciences (ARIES), Nainital, India. 
In this set up the entire chip covers a field of $\sim$13$\times$13 arcmin$^{2}$ on the sky. 
The read-out noise and gain of the CCD are 5.3e$^{-}$ and 10e$^{-}$/ADU, respectively. The observations were carried out in the 
binning mode of 2$\times$2 pixel to improve the signal-to-noise (S/N) ratio.  The full width at half-maximum (FWHM) of the star images 
were $\sim$2$-$3 arcsec. The sky flat frames and bias frames were also taken frequently during the observing runs. 
A number of short exposures in all the filters were also taken to avoid saturation of bright stars. The observations were standardized 
using the stars in the SA98 field \citep{1992AJ....104..340L} observed on 2006 January 23. The log of the observations is given in 
Table \ref{log}. To estimate the contamination due to the foreground/background field stars, a reference field of $\sim$13$\times$13 
arcmin$^{2}$ located at about 40$\arcmin$ away,  was also observed. 

The CCD data frames were reduced using the computing facilities available at ARIES, Nainital. 
Initial processing of the data frames was done using the {\small IRAF} (Image Reduction and Analysis Facility)\footnote{{\small IRAF} is distributed 
by the National Optical Astronomical Observatories, USA.} and {\small ESO-MIDAS} (European Southern Observatory Munich 
Image Data Analysis System)\footnote{{\small ESO-MIDAS is developed and maintained by the European
Southern Observatory.}} data reduction packages. The details of the data reduction can be found in our earlier works 
\citep[e.g.,][]{2007PASJ...59..547P,2008MNRAS.384.1675J}. 

To translate the instrumental magnitudes to the standard magnitudes, the following calibration equations, derived using a 
least-square linear regression, were used:

\noindent
\begin{small}
$u = U$ + (6.670 $\pm$ 0.011) + (0.044 $\pm$ 0.008)$(U-B)$ + (0.582 $\pm$ 0.013)$X$, \\
$b = B$ + (4.519 $\pm$ 0.006) + (-0.026 $\pm$ 0.004)$(B-V)$ + (0.332 $\pm$ 0.006)$X$, \\
$v = V$ + (4.096 $\pm$ 0.004) + (-0.029 $\pm$ 0.003)$(V-I)$ + (0.224 $\pm$ 0.004)$X$, \\
$r = R_{c}$ + (4.018 $\pm$ 0.006) + (-0.012 $\pm$ 0.007)$(V-R)$ + (0.165 $\pm$ 0.006)$X$, \\
$i = I_{c}$ +(4.559 $\pm$ 0.006) + (-0.055 $\pm$ 0.003)$(V-I)$ + (0.117 $\pm$ 0.006)$X$ \\
\end{small}
\noindent

where $U$, $B$, $V$, $R_{C}$ and $I_{C}$ are the standard magnitudes and $u$, $b$, $v$, $r$ and 
$i$ are the instrumental 
aperture magnitudes normalized for 1 second of exposure time and $X$ is the airmass. 
The second-order color correction terms were ignored as they are generally small 
in comparison to other errors present in the photometric data reduction. The standard deviations of 
the standardization residuals, $\Delta$, between standard and transformed $V$ magnitude and $(U-B)$, $(B-V)$, $(V-R)$ and $(V-I)$ 
colors of the standard stars are 0.02, 0.04, 0.02, 0.02 and 0.01 mag, respectively. The typical DAOPHOT errors in magnitude as a 
function of corresponding magnitude in different passbands are found to increase with the magnitude and become large 
($\geq$ 0.1 mag) for stars fainter than V $\simeq$ 21 mag. The measurements beyond this magnitude limit were not 
considered in the analysis. The photometric data along with positions of the stars are given in Table \ref{phot_data}. 
A sample and format of the Table is shown here. The complete table is available only in electronic form as a part of the online material.

\begin{table*}
\centering
\scriptsize
\caption{$UBV(RI)_{C}$ photometric data. The complete table is available in the electronic form.}
\label{phot_data}
\begin{tabular}{cccccccc}\hline \hline
Star ID  & R.A  (J2000)   & DEC (J2000)  &  $U \pm \sigma_{U}$ &  $B \pm \sigma_{B}$  & $V \pm \sigma_{V}$ & $R_{C} \pm \sigma_{R_{C}}$  & $I_{C} \pm \sigma_{I_{C}}$  \\
   &   (h~m~s)  & ($\degr~\arcmin~\arcsec$) &  (mag) & (mag) & (mag) & (mag) & (mag)    \\
 (1)   &  (2)     &   (3)    &   (4)  &  (5)   &   (6)   &  (7)      &  (8)   \\
\hline
     1 &   5~30~59.954  & 34~15~31.95  &   13.085 $\pm$    0.007 &   12.063 $\pm$    0.005 &   10.759 $\pm$    0.006 &   10.050 $\pm$    0.008 &    9.398 $\pm$    0.013 \\ 
     2 &   5~31~57.547  & 34~10~47.45  &   11.520 $\pm$    0.007 &   11.449 $\pm$    0.008 &   10.872 $\pm$    0.009 &   10.520 $\pm$    0.002 &   10.161 $\pm$    0.009 \\ 
     3 &   5~31~26.307  & 34~11~9.94   &   11.036 $\pm$    0.005 &   11.411 $\pm$    0.009 &   11.136 $\pm$    0.010 &   10.918 $\pm$    0.005 &   10.643 $\pm$    0.005 \\ 
     4 &   5~31~40.582  & 34~10~52.00  &   11.694 $\pm$    0.006 &   11.599 $\pm$    0.009 &   11.193 $\pm$    0.011 &   10.948 $\pm$    0.005 &   10.676 $\pm$    0.007 \\ 
     5 &   5~31~40.243  & 34~14~26.25  &   11.966 $\pm$    0.008 &   11.643 $\pm$    0.006 &   11.252 $\pm$    0.012 &   11.017 $\pm$    0.006 &   10.718 $\pm$    0.010 \\ 
 -- &      --         &     ---          &    --     &    --      &   --  &   -- &   --  \\ 
 -- &      --         &     ---          &    --     &    --      &   --  &   -- &   --  \\ 
 -- &      --         &     ---          &    --     &    --      &   --  &   -- &   --  \\ 
\hline
\hline
\end{tabular}
\end{table*}

\subsubsection{Completeness of the data} \label{completeness}

The study of  the luminosity functions (LFs)/MFs requires necessary corrections in the data sample 
to take into account the incompleteness that may occur due to various reasons (e.g. crowding of the stars). 
We used the ADDSTAR routine of DAOPHOT II to determine the completeness factor (CF). 
The procedures have been outlined in detail in our earlier works \citep{2001A&A...374..504P,2005MNRAS.358.1290P}. 
The CF as a function of $V$ magnitude is given in Table \ref{tab_completeness}, 
which  indicates that present optical data have $\sim$ 90 per cent completeness at $V$ $\sim$ 20 mag. 
As expected the incompleteness increases with the magnitude. 

\begin{table}
\centering
\scriptsize
\caption{Completeness factor (CF) of the optical photometric data in the cluster and field regions.}
\label{tab_completeness}
\begin{tabular}{ccc} 
\hline
$V$ range &   Cluster region   & Field region \\
(mag)& $r\le3.5^\prime$  &   \\
\hline
 10-11   &1.00  &   1.00 \\
 11-12   &1.00  &   1.00 \\
 12-13   &1.00  &   1.00 \\
 13-14   &1.00  &   1.00 \\
 14-15   &1.00  &   1.00 \\
 15-16   &1.00  &   1.00 \\
 16-17   &1.00  &   0.98 \\
 17-18   &0.97  &   0.97 \\
 18-19   &0.94  &   0.96 \\
 19-20   &0.92  &   0.92 \\
 20-21   &0.88  &   0.86 \\
\hline
\end{tabular}
\end{table}

\subsubsection{Comparison with previous studies}

\begin{figure}
\centering
\resizebox{7.725cm}{8.725cm}{\includegraphics{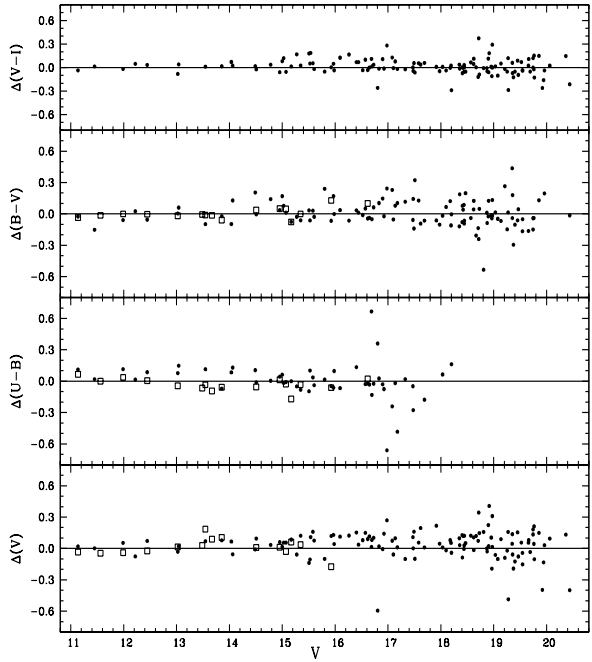}}
\caption{Comparison between the present photometric data with those available in the literature 
i.e., $\Delta$ (literature$-$present data) as a function of $V$ magnitude. The squares and circles represent 
the comparison of present data with those by \citet{1986Ap&SS.120..107P} and \citet{1994BASI...22..291B}, respectively.}
\label{fig_phot_comparison}
\end{figure}

A comparison of the present photometric data with those available in the literature has been carried out and 
the difference $\Delta$ (literature - present data) as a function of $V$ magnitude is plotted in Fig. \ref{fig_phot_comparison}.
The comparison indicates that the present $V$ mag and colors are in good agreement with the CCD and 
photoelectric photometry by \citet{1986Ap&SS.120..107P} and \citet{1994BASI...22..291B}, respectively.

\subsection{Polarimetric data}

Polarimetric observations were carried out on two nights (2010 November 12 and 2010 December 13), 
using the ARIES Imaging Polarimeter 
\citep*[AIMPOL;][]{2004BASI...32..159R} mounted at the Cassegrain focus of the 104-cm Sampurnanand 
telescope of the ARIES, Nainital, India. 
The details of the AIMPOL are given in our earlier works (E11 and E12).  
The observations were carried out in the $V$, $R_{c}$ and $I_{c}$ photometric bands. 
The detailed procedures used to estimate the polarization and position angles for the program stars are given in E11 and E12.

The instrumental polarization of the AIMPOL is estimated  to be less than 0.1 per cent in all the bands (E11, E12). 
The poarization measurements were corrected for both the null polarization
($\sim$ 0.1 per cent), which is independent of the passbands, and the zero-point polarization angle by 
observing several unpolarized and polarized standard stars from \citet*[][here after S92]{1992AJ....104.1563S}. 
The results for the polarized standard stars are given in Table \ref{stand_results}. 
The values of $\theta$ are in equatorial coordinate system measured eastwards from the
North. Both, the observed degree of polarization [$P(\%)$]
and polarization angle [$\theta (\degr)$] for
the polarized standards are
in good agreement with those given by S92.

\begin{table}
\centering
\scriptsize
\caption{Observed polarized standard stars}
\label{stand_results}
\begin{tabular}{lllll}\hline \hline
 & P$\pm\epsilon_{P}$($\%$) & $\theta\pm\epsilon_{\theta}$($\degr$) &P$\pm\epsilon_{P}$($\%$) &  
                             $\theta\pm\epsilon_{\theta}$($\degr$)\\
\multicolumn{5}{c}{\hspace{0.5cm}Our work\hspace{2.4cm} Schmidt et al. (2002)}\\
\hline
\multicolumn{5}{c}{Polarized standard stars}\\
\hline
\multicolumn{5}{c}{2010 November 12}\\
\hline
\multicolumn{5}{c}{{\bf HD\,19820}}\\
$B$     &   4.54 $\pm$ 0.11 &  116.3 $\pm$    0.7  &    4.70 $\pm$ 0.04  &  115.7  $\pm$   0.2  \\
$V$     &   4.74 $\pm$ 0.10 &  115.1 $\pm$    0.6  &    4.79 $\pm$ 0.03  &  114.9  $\pm$   0.2  \\
$R_{c}$ &   4.57 $\pm$ 0.07 &  114.4 $\pm$    0.4  &    4.53 $\pm$ 0.03  &  114.5  $\pm$   0.2  \\ 
$I_{c}$ &   3.97 $\pm$ 0.07 &  115.4 $\pm$    0.5  &    4.08 $\pm$ 0.02  &  114.5  $\pm$   0.2  \\ 
\multicolumn{5}{c}{{\bf HD\,204827}}\\ 
$B$     &  5.75 $\pm$ 0.14  &     57.6  $\pm$   0.7   &     5.65  $\pm$   0.02   &    58.2   $\pm$    0.1 \\ 
$V$     &  5.47 $\pm$ 0.10  &     63.3  $\pm$   0.5   &     5.32  $\pm$   0.01   &    58.7   $\pm$    0.1 \\ 
$R_{c}$ &  4.93 $\pm$ 0.09  &     59.2  $\pm$   0.5   &     4.89  $\pm$   0.03   &    59.1   $\pm$    0.2 \\ 
$I_{c}$ &  4.07 $\pm$ 0.09  &     59.1  $\pm$   0.6   &     4.19  $\pm$   0.03   &    59.9   $\pm$    0.2 \\
\hline
\multicolumn{5}{c}{2010 December 13}\\
\hline
\multicolumn{5}{c}{{\bf HD\,19820}}\\
$B$     &  4.47  $\pm$ 0.11   &     115.3  $\pm$    0.7   &      4.70   $\pm$  0.04  &    115.7  $\pm$    0.2 \\
$V$     &  4.78  $\pm$ 0.09   &     115.6  $\pm$    0.5   &      4.79   $\pm$  0.03  &    114.9  $\pm$    0.2 \\
$R_{c}$ &  4.60  $\pm$ 0.07   &     114.2  $\pm$    0.4   &      4.53   $\pm$  0.03  &    114.5  $\pm$    0.2 \\
$I_{c}$ &  3.99  $\pm$ 0.07   &     114.3  $\pm$    0.5   &      4.08   $\pm$  0.02  &    114.5  $\pm$    0.2 \\
\multicolumn{5}{c}{{\bf HD\,25443}}\\
$B$     &   5.17  $\pm$  0.11   &  134.7  $\pm$    0.6   &    5.23  $\pm$  0.09  &  134.3    $\pm$   0.5 \\
$V$     &   5.24  $\pm$  0.09   &  134.7  $\pm$    0.5   &    5.13  $\pm$  0.06  &  134.2    $\pm$   0.3 \\
$R_{c}$ &   4.97  $\pm$  0.09   &  134.3  $\pm$    0.5   &    4.73  $\pm$  0.05  &  133.6    $\pm$   0.3 \\
$I_{c}$ &   4.27  $\pm$  0.10   &  134.6  $\pm$    0.6   &    4.25  $\pm$  0.04  &  134.2    $\pm$   0.3 \\
\hline \hline
\end{tabular}
\end{table}

\section{Archival data} \label{archival_data}

\subsection{2MASS near-infrared data}\label{2mass_data}

NIR $JHKs$ data for point sources in the NGC\,1931 region have been obtained from the Two Micron 
All Sky Survey 
~\citep[2MASS;][]{2003yCat.2246....0C}. 
The 2MASS data reported to be 99\% complete 
up to $\sim$ 15.7, 15.1, 14.3 mag in $J$, $H$, $K_{s}$ bands, 
respectively\footnote{See http://www.ipac.caltech.edu/2mass/releases/allsky/\\
doc/sec6\_5a1.html}. 
To ensure photometric accuracy, we used only those photometric data which have quality flag ph-qual=AAA, which endorses 
a S/N $\geq$ 10 and photometric uncertainty $\textless$ 0.10 mag. The NIR data are used to identify the 
classical T-Tauri stars (CTTSs) and weak line T-Tauri stars (WTTSs).

\subsection{Spitzer IRAC data}

The archived MIR data observed with the {\it Spitzer} Infrared Array Camera (IRAC) have also been used in the present study. 
We obtained basic calibrated data (BCD) using the software Leopard. The exposure time of each BCD was 
10.4~sec and for each mosaic, 169 and 139 BCDs respectively, in Ch1 (3.6 $\mu$m) and Ch2 (4.5 $\mu$m) 
have been used. Mosaicking was performed using the 
MOPEX software provided by Spitzer Science Center (SSC). All of the mosaics were built at the native 
instrument resolution of 1.2 arcsec pixel$^{-1}$ with the standard BCDs. In order to avoid source 
confusion due to crowding, PSF photometry for all the sources was carried out using the DAOPHOT 
package available with the IRAF photometry routine. The detections are also examined visually in each 
band to remove non-stellar objects or false detections. The sources with photometric uncertainties 
$\textless$ 0.2 mag in each band were considered as good detections. A total of 3950 and 2793 sources were detected in 
the 3.6 and 4.5 $\mu$m bands. Aperture photometry for well isolated sources was done using an aperture 
radius of 3.6 arcsec with a concentric sky annulus of the inner and outer radii of 3.6 and 8.4 arcsec, respectively. 
For a standard aperture radius (12 arcsec) and background annulus of 12$-$22.4 arcsec, we adopted 
zero-point magnitude as 19.670 and 18.921 for the 3.6 and 4.5 $\mu$m bands, respectively. 
Aperture corrections were also made using the values described in 
the IRAC Data Handbook (Reach et al. 2006). 
The necessary aperture correction for the PSF photometry was then calculated from the selected isolated 
sources and were applied to the PSF magnitudes of all the sources. 

\section{STRUCTURE OF THE CLUSTER}\label{structure}

\citet{2004AJ....128.2306C} and \citet{2006AJ....132.1669S} have found that the initial stellar distribution in star 
clusters may be governed by the structure of the parental molecular cloud as well as 
how star formation proceeds in the cloud. Later evolution of the cluster may be governed by internal 
gravitational interaction among member stars and by external 
tidal forces due to the Galactic disk or giant molecular clouds.

\begin{figure*}
\centering
\resizebox{16cm}{15cm}{\includegraphics{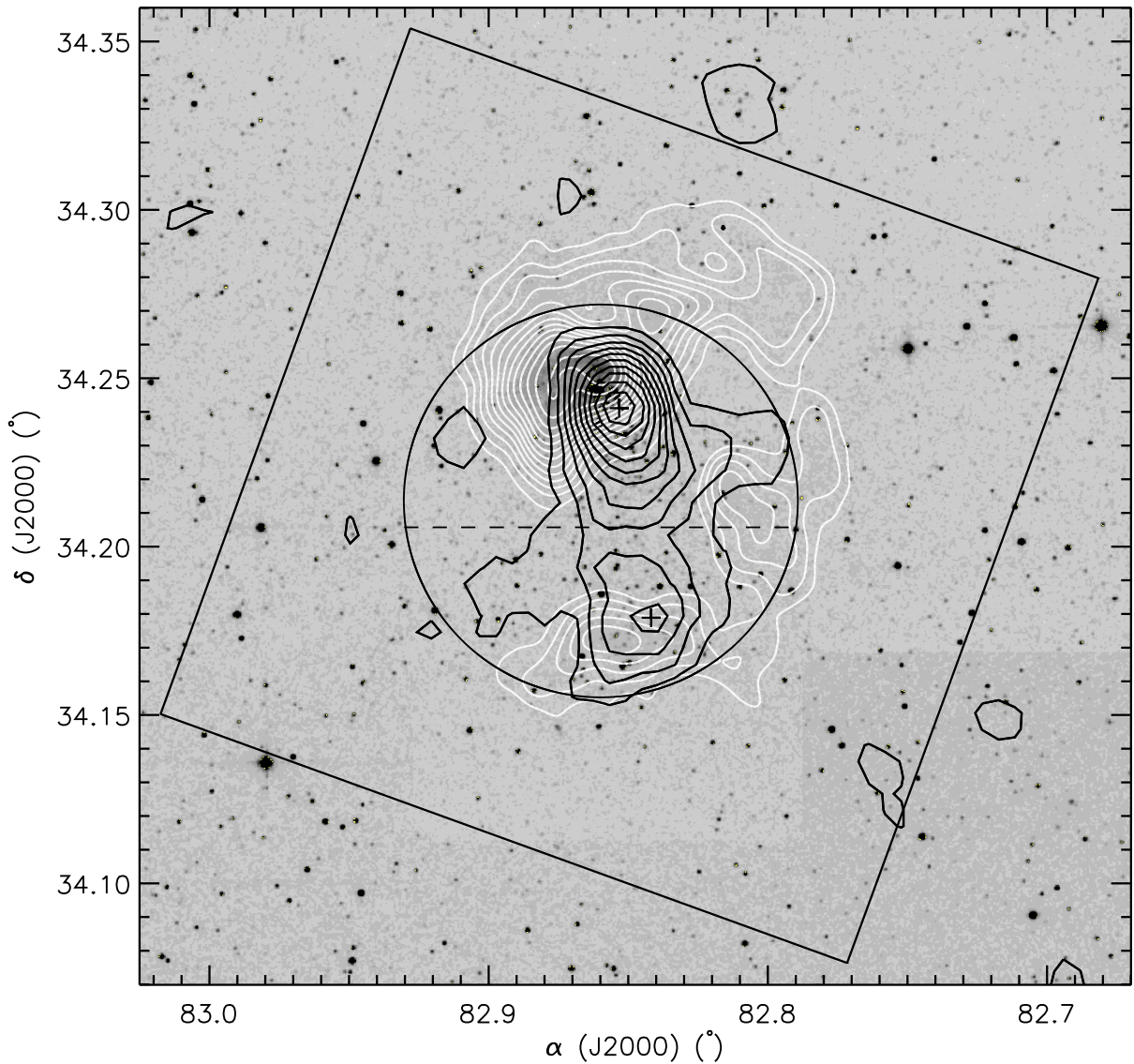}}
\caption{The isodensity contours (thick black curves) generated from the 2MASS 
$K_{s}$-band data (with $\sigma_{K_{s}}\textless$ 0.1 mag) using a grid size of 
$\simeq$35 $\times$ 35 arcsec$^2$. The contours are plotted above 1$\sigma$ level. The contours have a step size 
of 2 stars/arcmin$^2$ with the lowest contour representing 5 stars/arcmin$^2$. Isodensity contours 
manifest two prominent clusterings. The dashed line demarcates northern and southern parts of the region. 
The thick white contours represent the MSX A-band (8.3 $\mu$m) intensity distribution. The minimum and 
maximum contour levels of the MSX A-band intensity are 1.78$\times$10$^{-6}$ W/m$^{2}$-Sr and 3.35$\times$10$^{-5}$ W/m$^{2}$-Sr, 
respectively. The area marked with the continuous line represents the region covered by optical photometry. The circle represents 
the estimated boundary of the region. The plus sign represents the center of the clusters.}
\label{stellar_density_contours}
\end{figure*}

The isodensity contours shown in Fig. \ref{stellar_density_contours}, 
obtained using the stars detected in the 2MASS $K_{s}$-band ($\sigma_{Ks} \textless$ 0.1 mag), 
are used to study the morphology of the cluster. The contours are plotted above 1$\sigma$ level. The surface density distribution reveals  
two prominent structures distributed around $\alpha$(2000)=05$^{h}$31$^{m}$24$\fs$781, 
$\delta$(2000)=+34$\degr$14$\arcmin$28$\farcs$05 and $\alpha$(2000)=05$^{h}$31$^{m}$22$\fs$034, 
$\delta$(2000)=+34$\degr$10$\arcmin$43$\farcs$99, suggesting the presence  of a double cluster in the region. 
In fact the radial density profile (RDP) of the region by \citet{2009MNRAS.397.1915B} also reveals a density 
enhancement around the radial distance of $\sim$ 3$-$5 arcmin. 

\begin{figure}
\centering
\resizebox{8.275cm}{8.275cm}{\includegraphics{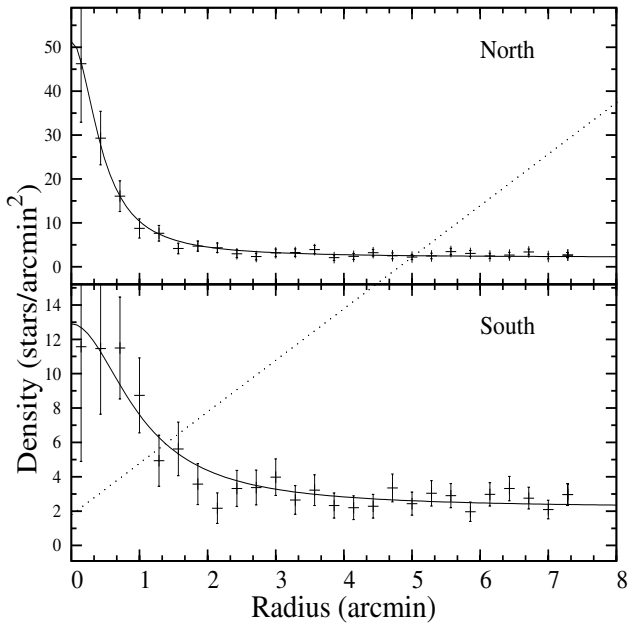}}
\caption{The radial density profile (RDP) is derived using the 2MASS $K_s$-band data (using $\sigma_{K_{s}}~\textless$ 0.1) by
dividing star counts inside the concentric annulus of 30 arcsec width around the cluster center by the respective area.
The densities thus obtained are plotted as a function of radius. The upper and lower panels show the RDPs for the
northern and southern clusters respectively.}
\label{rdp}
\end{figure}

We used the star count technique to estimate the radial extent of the two clusters. 
The points of maximum densities in Fig. \ref{stellar_density_contours} were considered as the center of the clusters. 
The RDP is derived using the 2MASS $K_s$-band data ($\sigma_{K_{s}}~\textless$ 0.1 mag) by dividing star counts 
inside the concentric annulus of 30 arcsec width around the cluster center by the respective area. 
The densities thus obtained are plotted as a function of radius 
in Fig. \ref{rdp}, where, 1 arcmin at the distance of the cluster (2.3 kpc, cf. Section \ref{dist_age}) 
corresponds to $\sim$ 0.67 pc. The upper and lower panels show the RDPs for the northern and southern clusters, respectively.  
The error bars are derived assuming that the number of stars in each annulus follows Poisson statistics.

The radial extent of the clusters ($r_{cl}$) is defined as the point where the cluster stellar density merges with 
the field stellar density. Within the errors, the observed RDPs for both the clusters seem to merge 
with the background field at a radial distance of $\sim$ 2 arcmin. 
Hence, we assume a radius of 2 arcmin for both the clusters. 

\section{ANALYSIS OF THE POLARIMETRIC DATA} \label{analysis}

\subsection{Polarization vector map \& Distribution of $P_{V/R/I}$~$(\%)$ and $\theta_{V/R/I}$~$(\degr)$}\label{distri_p_t}

\begin{figure*}
\centering
\resizebox{16cm}{13cm}{\includegraphics{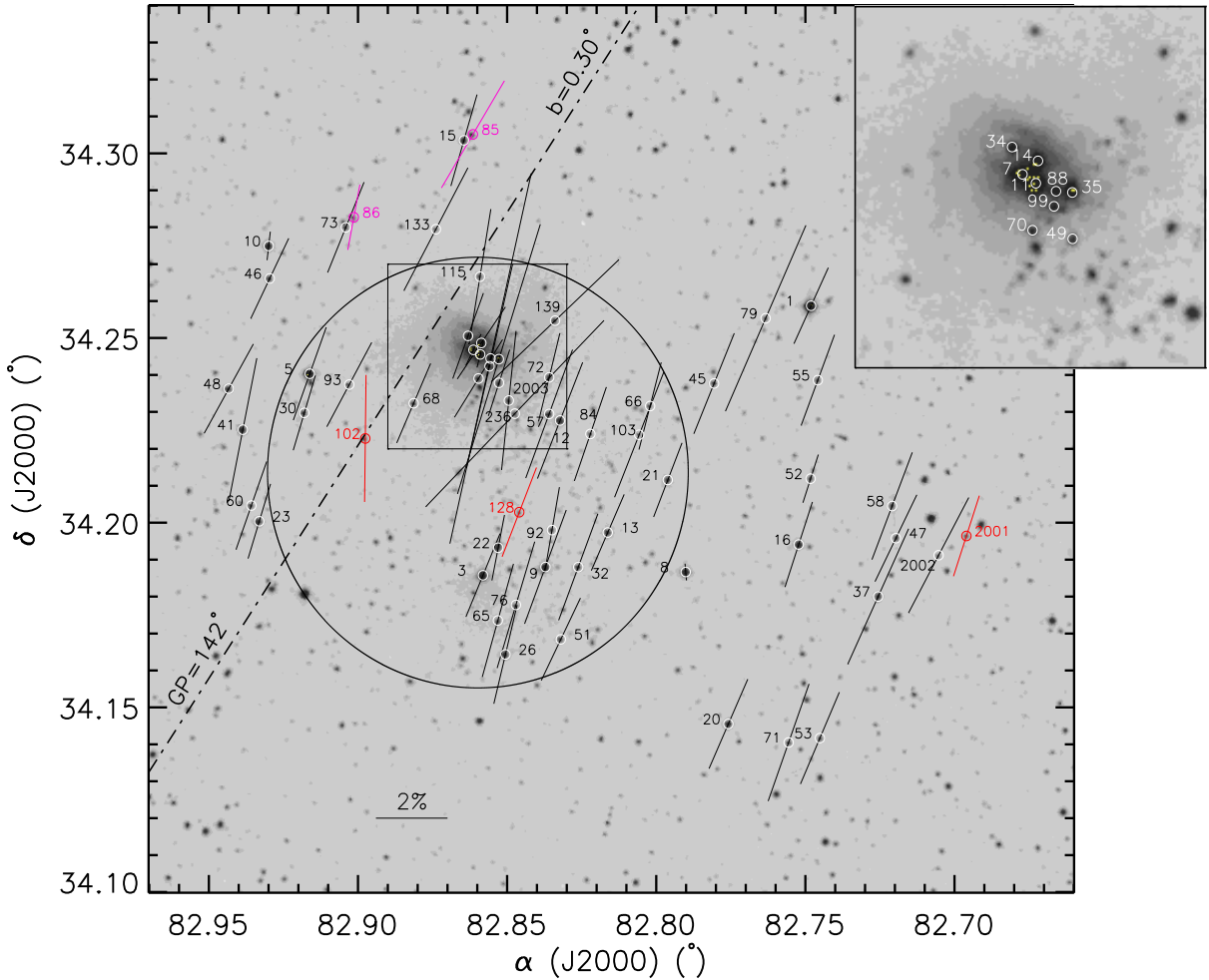}}
\caption{The polarization measurements of the stars in NGC\,1931 are superimposed on the 2MASS $K_{s}$-band image. 
The stars with $V$-band polarimetric measurements are shown with black vectors. 
The red and magenta vectors represent polarimetric measurements in $R_{C}$ and $I_{C}$-bands. 
The dashed-dotted line represents the orientation of the projection of the GP at $b=0\fdg30$ which 
corresponds to a position angle of 142$\degr$. 
A vector with a 2 per cent of polarization at a position angle of 90$\degr$ is 
drawn for a reference. Barring a few vectors near the cluster center (shown in a square box), 
the majority of the vectors are closely aligned with GP indicating an ordered magnetic field towards NGC\,1931. The 
vectors in the nebulous region of the cluster show a scattered distribution. 
Two stars \#139 and \#236 show high degree of polarization as well as different polarization angles 
from the rest of the stars. These stars are identified as probable YSOs (cf. Sec \ref{few_ysos}). 
The box at the corner shows an enlarged view of the central region. See the electronic version for the color 
image.} 
\label{pol_vectors}
\end{figure*}
Table \ref{VRI_poldata} lists the polarization measurements for 62 stars towards the cluster region. 
The star identification numbers for 59 stars are taken from column 1 of Table \ref{phot_data}, 
whereas three stars \#2001, \#2002 and \#2003, do not have optical data. 
The degree of polarization P (in per cent), polarization angles 
$\theta$ (in degree) measured in $V, (R, I)_{c}$ bands and their corresponding standard errors 
are given in columns $4-9$. The black vectors (Fig. \ref{pol_vectors}) show the sky projection of 
the $V$-band polarization measurements. 
In the case of a few stars for which the $V$-band polarization measurements are not available, 
we plotted the polarization measurements either of $R_{c}$-band (red) or $I_{c}$-band (magenta). 
The length of each polarization vector is proportional to the degree of polarization. 
The dashed-dotted line represents the orientation of the 
projection of the Galactic plane ($GP$) at $b$ = 0.30$\degr$, which corresponds to a position angle of 142$\degr$. 
Interestingly, all the polarization vectors except those near the cluster center (enclosed by a square box) are roughly 
parallel to the GP. 
The enlarged view of the central region is shown in a separate panel at the top right part of the figure. 
The polarization measurements of two stars \#7 and \#11 could be
affected by their close companions as well as by the nebulosity. 
Hence, the polarization measurements of these stars could be less reliable. 

\begin{figure}
\centering
\resizebox{8.275cm}{8.275cm}{\includegraphics{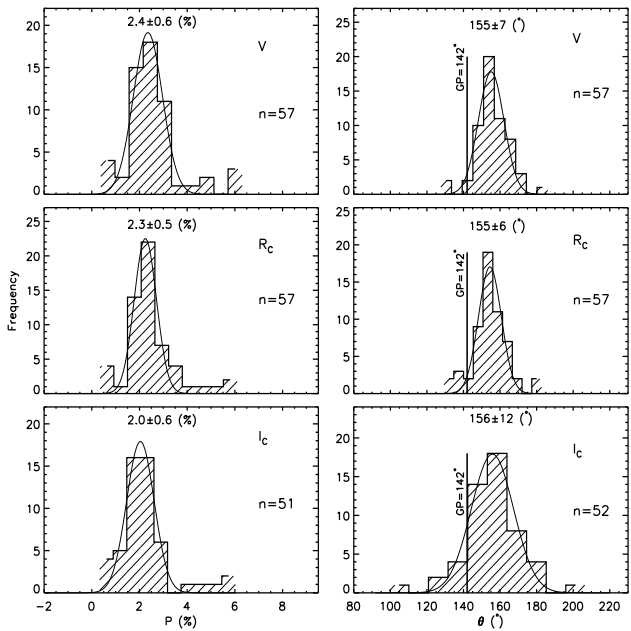}}
\caption{Frequency distribution  of polarization (left panel) and polarization angle values (right panel) 
for the stars detected in $V, (R, I)_{c}$ bands. In each case, the mean and standard deviation of polarization and 
polarization angle values are obtained by fitting a Gaussian function. The number of sources detected in 
each passband is also indicated. The position angle corresponds to the $b=0.30\degr$ (GP) is also shown at 
142$\degr$ with a thick line. The mean polarization angles differ from the GP by $\sim$13$\degr$.}
\label{gaussfit_p_pa}
\end{figure}

The distribution of $P$ and $\theta$ in $V, (R, I)_{c}$ bands is shown in 
Fig. \ref{gaussfit_p_pa}. A Gaussian fit to the each distribution yields a mean and a standard deviation as 
$P_V=2.4\pm$0.6 per cent, $P_R=2.2\pm$0.5 per cent, $P_I=2.0\pm$0.6 per cent, and  
$\theta_V=155\pm7\degr$, $\theta_R=154\pm6\degr$, 
$\theta_I=156\pm12\degr$. The values of $P_V (2.4\pm0.6$ per cent) and 
$\theta_V (155\pm7\degr)$ towards NGC\,1931 are comparable 
to those ($P_V=2.3\pm$0.1 per cent and $\theta_V=160\pm3\degr$, E11) for  
Stock\,8 (${\it l}=173.37\degr$, $b=-0.18\degr$, distance = 2.05 kpc) which is spatically located near the NGC\,1931. 

\subsection{Member identification}\label{membership} 

The identification of probable members of a cluster and the foreground/background stars towards the direction of 
the cluster is necessary to study the nature of the dust properties of the ICM/ISM and the magnetic field 
associated with the foreground and the ICM.  Our earlier studies (E11 \& E12) have shown that the polarimetry in combination 
with the $(U-B)-(B-V)$ two-color diagram (TCD) can be efficiently use to identify the probable members in the cluster region. 
This section describes the determination of membership using the polarization properties in combination with the  $(U-B)-(B-V)$ TCD. 

\begin{figure*}
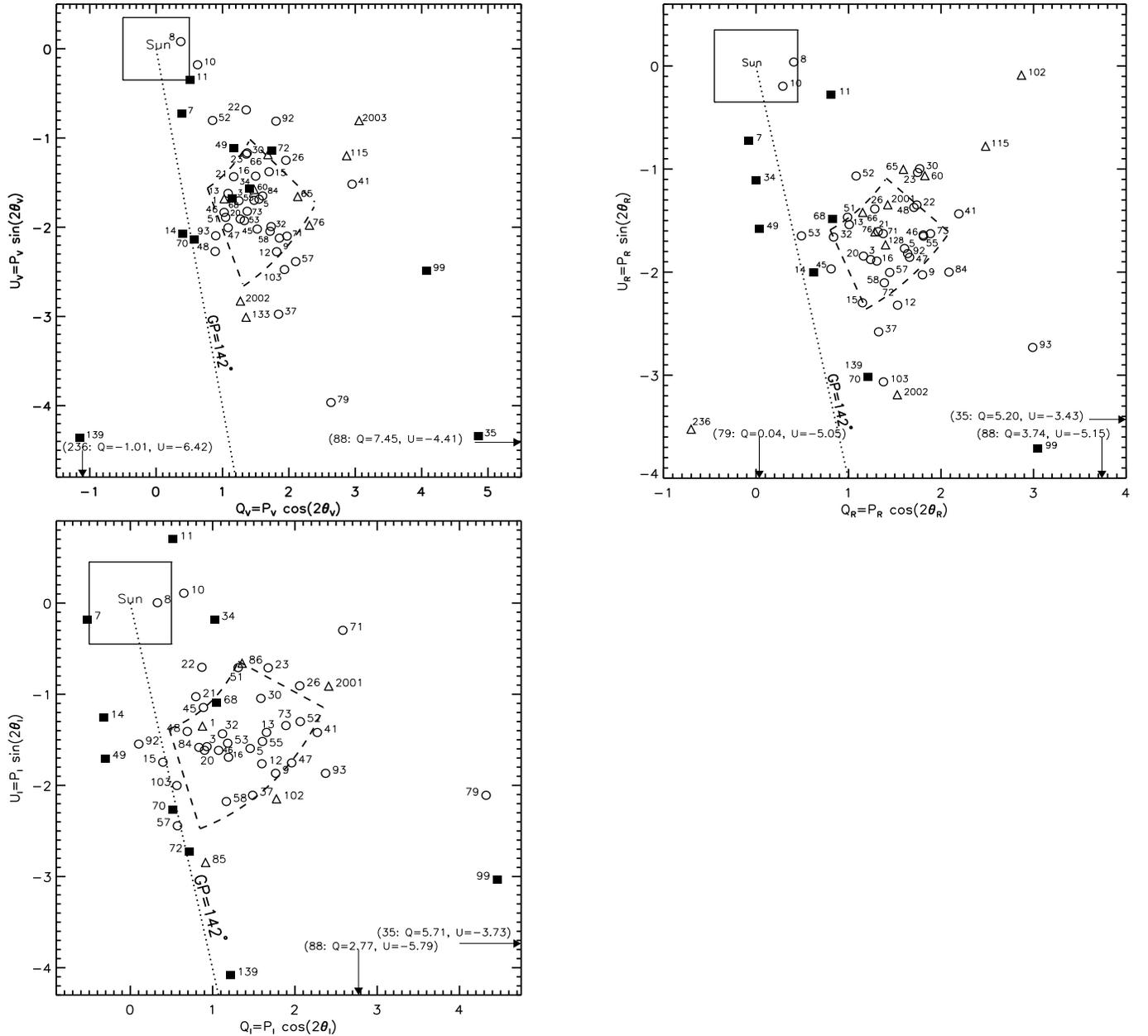

\resizebox{8.275cm}{8.275cm}{\includegraphics{Q_U_Stokesplane_extr_stars_NGC1931_V_band.epsi}}
\resizebox{8.275cm}{8.275cm}{\includegraphics{Q_U_Stokesplane_extr_stars_NGC1931_R_band.epsi}}
\resizebox{8.275cm}{8.275cm}{\includegraphics{Q_U_Stokesplane_extr_stars_NGC1931_I_band.epsi}}
\caption{$U$ versus $Q$ diagrams for all the stars given in Table \ref{VRI_poldata}. 
Open circles represent the stars having $V$-band data. The filled squares represent the stars 
located within the box as shown in Fig \ref{pol_vectors}. The triangles represent the stars 
having either single or double band polarimetric data. 
The box with dashed line marks the boundary of mean $P\pm\sigma$ and mean $\theta\pm\sigma$. 
The position of the Sun is also shown with a square box at the ($Q=0$, $U=0$) coordinates. The GP is drawn with a dotted line. 
The stars distributed within or nearby 1$\sigma$ box of all the three Stokes planes are considered as 
probable members associated with the cluster.}
\label{stokesplane_VRI}
\end{figure*}

The individual Stokes parameters of the polarization vectors of the V/R/I-band, $P_{V/R_{c}/I_{c}}$, given by 
$Q_{V/R_{c}/I_{c}}$ = $P_{V/R_{c}/I_{c}}$ $\cos$(2$\theta_{V/R_{c}/I_{c}}$) and 
$U_{V/R_{c}/I_{c}}$ = $P_{V/R_{c}/I_{c}}$ $\sin$(2$\theta_{V/R_{c}/I_{c}}$) have been estimated and 
are presented in the $U_{V/R_{c}/I_{c}}$ versus $Q_{V/R_{c}/I_{c}}$  plot as shown in Fig. \ref{stokesplane_VRI}. 
Since the polarization of light from a star depends on the column density of aligned dust 
grains that lie in front of the star, the cluster  members are expected to group together in the $U$ versus $Q$ plot, 
whereas non-members are expected to show a scattered distribution. Similarly, the polarization angles of cluster members 
would be similar but could be different for foreground or background non-member stars as light from them could be 
affected in a different manner because of contributions from different/additional dust components. 
Therefore, the $U-Q$ plot is a useful tool to segregate the members from non-members in the cluster region. 
However, the stars with intrinsic polarization, e.g., due to an asymmetric distribution of matter around 
YSOs and/or rotation in their polarization angles (see e.g., E11 \& E12), may also create scattered distribution in the $U-Q$ plane.  

The probable members of the cluster should have a similar location in all the three $U-Q$ diagrams. 
The Stokes parameters for the stars lying within the square box as shown in Fig. \ref{pol_vectors} and 
embedded in the nebulosity are shown with filled squares. 
The box with dashed line in $U-Q$ plots shows the boundary of 
the area having mean P $\pm$ $\sigma$  
($P_V$=2.4$\pm$0.6 per cent, $P_R$=2.3$\pm$0.5 per cent, 
$P_I$=2.0$\pm$0.6 per cent) and mean $\theta$ $\pm$ $\sigma$ 
($\theta_V$=155$\degr\pm7\degr$, $\theta_R$=155$\degr\pm6\degr$, 
$\theta_I$=156$\degr\pm12\degr$) obtained from the distribution shown in Fig. \ref{gaussfit_p_pa}. 
The stars lying within or near the 1$\sigma$ box in all the three $U-Q$ plots may be 
probable members associated with the cluster.

\begin{figure}
\centering
\resizebox{8.275cm}{8.275cm}{\includegraphics{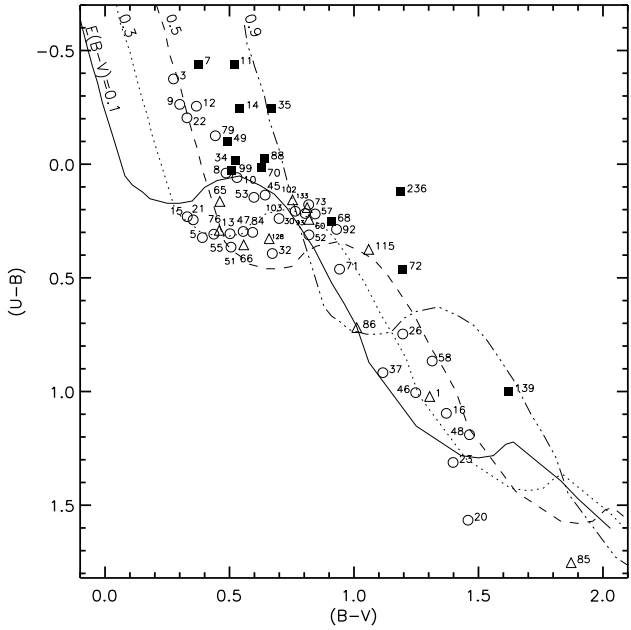}}
\caption{$(U-B)$ versus $(B-V)$ TCD  of the stars with polarimetric data. The symbols 
are the same as in the Fig. \ref{stokesplane_VRI}. The ZAMS 
from Schmidt-Kaler (1982) is shifted along a normal reddening vector having a slope of $E(U-B)/E(B-V)$ = 0.72. 
The TCD shows a variable reddening in the cluster region with $E(B-V)_{min}$ $\sim$ 0.5 mag and $E(B-V)_{max}$ $\sim$ 0.9 mag.  
}
\label{UBBVCCD}
\end{figure}

It is expected that the  members of a cluster should exhibit comparable $E(B-V)$ values, 
whereas the foreground (background) population is expected to be less (highly) extincted as compared to the 
cluster members. Fig. \ref{UBBVCCD} shows the 
$(U-B)-(B-V)$ TCD for only 58 stars as $U-B$ color is not available for three stars. 
The symbols of the stars are same as in Fig. \ref{stokesplane_VRI}. 
All the stars shown in Fig. \ref{UBBVCCD} have the 
polarimetric data. In Fig. \ref{UBBVCCD}, the zero-age main sequence (ZAMS) 
from Schmidt-Kaler (1982) is shifted along a normal reddening vector having a slope of $E(U-B)/E(B-V)$ = 0.72. The TCD shows 
a variable reddening in the cluster region with $E(B-V)_{min}$ $\sim$ 0.5 mag and $E(B-V)_{max}$ $\sim$ 0.9 mag. 
It is also apparent from Fig. \ref{UBBVCCD} that sources located 
in the northern nebulous region (lying within the box in Fig. \ref{pol_vectors}) 
reveals presence of parent molecular cloud and show a 
variable extinction of $E(B-V)$ $\sim$ $0.5-0.9$ mag. The polarimetric observations in combination with the 
$(U-B)-(B-V)$ $TCD$ are used to identify members of the NGC\,1931 cluster (cf. E11 and E12). 
The probable members thus identified are given in Table \ref{pol_ccd_members}.
The stars \#7, 11, 14, 35, 49, 88 and 99 are embedded in the northern nebulosity, hence 
the polarization measurements may be affected by the nebulosity, consequently may show scattered distribution 
in the $U$ vs $Q$ diagram. Out of 22 probable members, 11 stars (\#3, 7, 11, 14, 21, 22, 35, 49, 79, 88 and 93) 
have either $\sigma_{1} \textgreater$ 1.5 and/or $\overline\epsilon \textgreater$ 2.3, indicating for 
the presence of intrinsic polarization and/or rotation in their polarization angles (cf. Sec \ref{dust_properties}). 

\begin{table}
\centering
\scriptsize
\caption{The list of probable members of the cluster selected on the basis of polarimetric and photometric observations.}
\label{pol_ccd_members}
\begin{tabular}{ccc}\hline \hline
Star ID  & $E(B-V)$ & lying within 3\farcm5 region \\
   &  (mag)   &  \\
\hline
\hline
   3  &   0.47  & yes \\
   5  &   0.38  & yes$^\star$ \\
   7  &   0.61  & yes$^\dagger$ \\
   9  &   0.46  & yes \\
  11  &   0.79  & yes$^\dagger$ \\
  12  &   0.54  & yes \\
  13  &   0.52  & yes \\
  14  &   0.75  & yes$^\dagger$ \\
  21  &   0.36  & yes$^\star$ \\
  22  &   0.48  & yes \\
  32  &   0.70  & yes \\
  35  &   0.91  & yes$^\dagger$ \\
  45  &   0.75  & no \\
  47  &   0.59  & no \\
  49  &   0.64  & yes$^\dagger$ \\
  51  &   0.51  & yes \\
  53  &   0.69  & no \\
  55  &   0.44  & no \\
  79  &   0.59  & no \\
  88  &   0.80  & yes$^\dagger$ \\
  93  &   0.92  & yes \\
  99  &   0.62  & yes$^\dagger$ \\
\hline
\hline
\end{tabular}\\
$\star$: The polarization of these sources is comparable to the cluster members, however 
the $E(B-V)$ values for these stars are less than minimum reddening ($E(B-V)_{min}$=0.50 mag) for the 
cluster region. \\
$\dagger$: These stars are embedded in the northern nebulosity. 
\end{table}

\subsection{Dust distribution}

\begin{figure}
\centering
\resizebox{8.275cm}{8.775cm}{\includegraphics{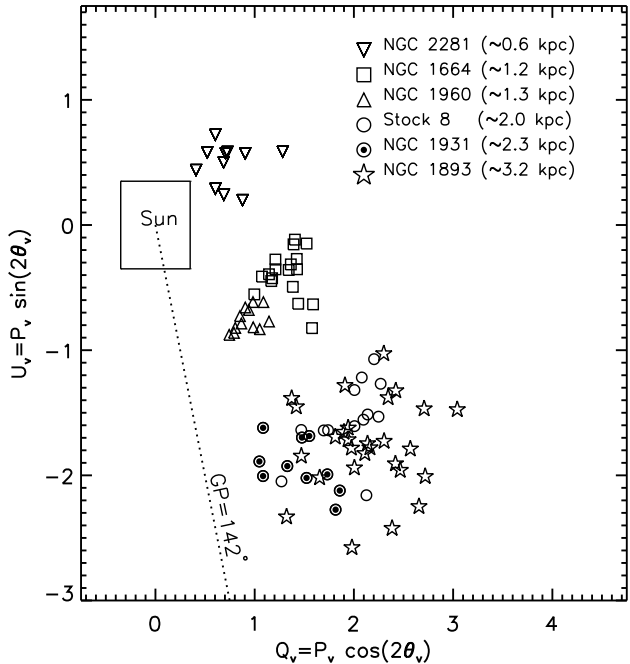}}
\caption{$U_{V}$ versus $Q_{V}$ plot of the 10 probable cluster members of NGC\,1931, which are free from 
possible intrinsic polarization and polarization angle rotation and also free from background nebulosity, 
along with other clusters towards 
the direction of NGC\,1931. The data for other clusters have been taken from E11.}
\label{stokesplane_all}
\end{figure}

In our previous studies (E11 and E12) we have  shown that the Stokes plane can be used to study the distribution of 
dust layers, the role of dust layers in polarization, the associated magnetic field orientation, etc. 
The vector connecting two points in the Stokes plane represents the amount of polarization, whereas the change in the 
direction of vectors indicates a change in the polarization angle. In the case of uniformly 
aligned dust grains (i.e. uniform magnetic field orientation), the degree of polarization is expected to increase with distance, 
but the direction of polarization (polarization angle) should remain the same and hence the Stokes vector should not change 
its direction with increasing distance. For example, in the case of NGC\,1893 (E11), 
the degree of polarization was found to increase with distance, whereas the direction of polarization remained almost 
constant (cf. their figure 5). In contrast, the magnetic field orientation shows systematic rotation 
with distance towards the direction of Berkeley\,59 (E12, see their figure 7) as the radiation from the cluster members might have 
a depolarization effect due to the systematic change in the dust grain alignment in the foreground medium. NGC\,1931 is located towards the 
direction of NGC\,1893 hence a similar behavior is expected for the cluster NGC\,1931 also. 
Fig. \ref{stokesplane_all} shows the $U_V-Q_V$ distribution for probable members of NGC\,1931 along with the data for other 
clusters NGC\,2281 ($\sim$ 0.6 kpc), NGC\,1664 ($\sim$ 1.2 kpc), NGC\,1960 ($\sim$ 1.3 kpc), 
Stock\,8 ($\sim$ 2.0 kpc), NGC\,1893 ($\sim$ 3.2 kpc), located towards the direction of NGC\,1931. The data for 
other clusters have been taken from E11. Out of 22 probable members (cf. Table \ref{pol_ccd_members}) of NGC\,1931, 
we have used only 10 stars (\#5, 9, 12, 13, 32, 45, 47, 51, 53 and 55) in Fig. \ref{stokesplane_all} 
which are free from intrinsic polarization and/or rotation in their polarization angles and also free from the background 
nebulosity. The polarization data for NGC\,1931 is consistent 
with the fact that the degree of polarization increases with the column density of dust grains 
lying in front of the stars that are relatively well aligned. 

The various dust layers located between the star and the observer can cause a sudden increase in degree of polarization.  
The number of such sudden increase has been used by E11 to characterize the dust layers encountered by the radiation 
along its path and the relative magnetic field orientations of the dust layers towards the direction of NGC\,1893 which is 
spatially located near the cluster NGC\,1931. The polarimetric observations for NGC\,1931 obtained in the present study 
follow the general trend revealed by the stars and clusters located towards the anticenter direction of the Galaxy as described by E11. 

\subsection{Dust properties}\label{dust_properties}

The wavelength dependence of polarization in the Galaxy can be represented by the 
following relation \citep*{1973IAUS...52..145S,1974AJ.....79..581C,1982AJ.....87..695W}; 
\begin{equation}\label{serkeq}
P_{\lambda} = P_{max} ~\exp[-K ~ln^{2} (\lambda_{max}/\lambda)] \\
\end{equation}
\noindent
where $P_{\lambda}$ and $\lambda_{max}$ are the percentage polarization at wavelength $\lambda$ and the peak polarization, 
occurring at wavelength  $\lambda_{max}$. The  $\lambda_{max}$ depends on the optical properties and characteristic 
particle size distribution of aligned grains \citep*{1975ApJ...196..261S,1978ApJ...225..880M} whereas the value of 
$P_{max}$ is dictated by the chemical composition, shape, size, column density, and alignment efficiency of the dust grains. 
The Serkowski’s relation with $K$=1.15 provides a reasonable representation of the observations of 
interstellar polarization between wavelengths 0.36 and 1.0~$\mu$m. The $P_{max}$ and $\lambda_{max}$ are obtained using the weighted 
least-squares fitting to the measured polarization in $V(RI)_{c}$ bands to equation \ref{serkeq} by adopting $K$=1.15. 
The parameters $\sigma_{1}$ (the unit weight error of the fit for each star){\footnote{The values 
of $\sigma_{1}$ for each star are computed using the 
expression $\sigma_{1}^{2}=\sum(r_{\lambda}/\epsilon_{p\lambda})^{2}/(m-2)$; where $m$ is the 
number of colors and $r_{\lambda}=P_{\lambda} ~­P_{max} \exp[-K~ln^{2} (\lambda_{max}/\lambda)$}}, 
which quantifies the departure of the data from the standard Serkowski's law, and $\overline\epsilon$, 
the dispersion of the polarization angle for each star normalized by the 
average of the polarization angle errors \citep[cf. ][]{1993AJ....105..258M} 
were also estimated. The estimated values of $P_{max}$,  $\lambda_{max}$, $\sigma_{1}$ and $\overline\epsilon$ 
for each star are given in Table \ref{VRI_poldata}. 

\begin{figure}
\centering
\resizebox{8.275cm}{14.725cm}{\includegraphics{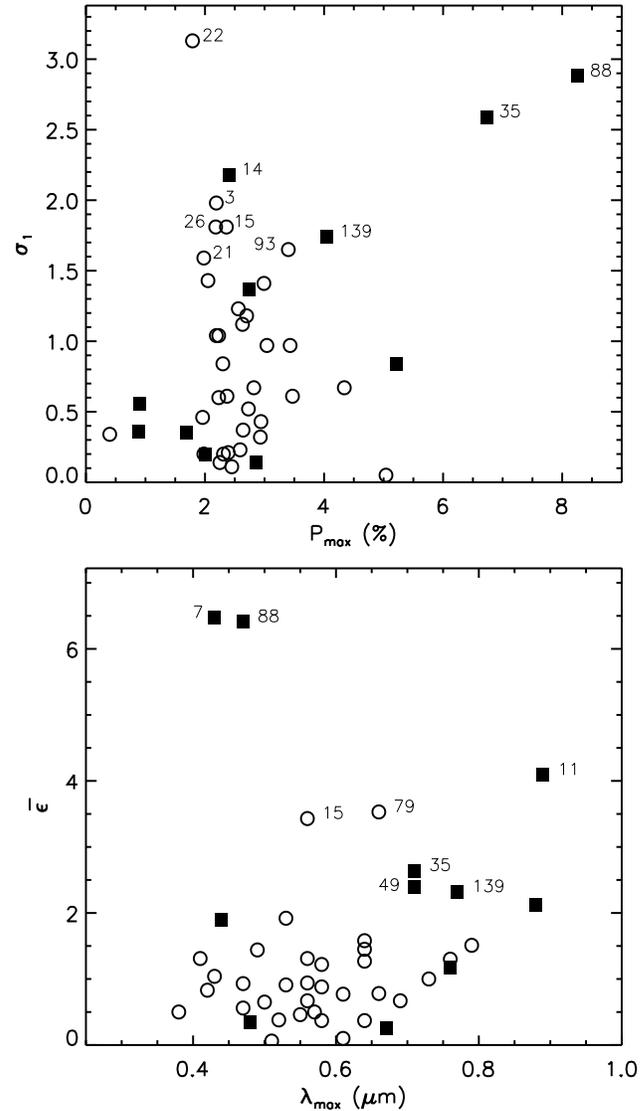}}
\caption{({\it upper panel}): $\sigma_{1}$ versus $P_{max}$ and ({\it lower panel}): 
$\overline\epsilon$ versus $\lambda_{max}$. Stars with $\sigma_{1}\textgreater$ 1.5, $\overline\epsilon$ 
$\textgreater$ 2.3 and with $\lambda_{max}$ $\textless$ 0.35$\mu$m (\#34) are marked with their IDs. The symbols are 
same as in Fig. \ref{stokesplane_VRI}}
\label{sig_vs_pmax_and_eps_vs_lmax}
\end{figure}

Fig. \ref{sig_vs_pmax_and_eps_vs_lmax} shows $\sigma_{1}$ versus $P_{max}$ (upper panel) and $\overline\epsilon$ versus 
$\lambda_{max}$ (lower panel) plots. The criteria mentioned above indicate that majority of the stars do 
not show evidence of intrinsic polarization. However, a few stars (15) show indication of either intrinsic polarization 
and/or rotation in their polarization angles. Ten stars (\#3,  14, 15, 21, 22, 26, 35, 88, 93 and 139) are found to 
exhibit an indication of intrinsic polarization 
as they have $\sigma_{1}$ $\textgreater$ 1.5 and 
8 stars (7, 11, 15, 35, 49, 79, 88 and 139) are found to show rotation in their polarization angles, whose 
$\overline\epsilon$ values are higher ($\textgreater$ 2.3) than that for the rest of the stars. 
Four stars (15, 35, 88 and 139) are found to show both indication of intrinsic polarization and also rotation 
in their polarization angles. 

The weighted mean values of $P_{max}$ and $\lambda_{max}$ using all the 22 probable members are found to be
2.51$\pm$ 0.03 per cent and 0.57$\pm$ 0.01 $\mu$m, respectively. 
Exclusion of stars showing intrinsic polarization does not change 
the mean values of $P_{max}$ and $\lambda_{max}$ significantly. The estimated $\lambda_{max}$ is slightly higher than the
value corresponding to the diffuse ISM ~\citep[0.545 $\mu$m; ][]{1975ApJ...196..261S}.
Using the relation $R_{V}$=(5.6$\pm$0.3)$\times$ $\lambda_{max}$ ~\citep{1978A&A....66...57W},
the value of $R_V$, the total-to-selective extinction, comes out to be 3.20 $\pm$ 0.05. 

However the $\lambda_{max}$ value (0.55$\pm$0.01$\mu$m; E11)
towards the cluster NGC\,1893, which is spatially close to NGC\,1931, is found to be
comparable to the value for the diffuse ISM (0.545$\mu$m; \citealt{1975ApJ...196..261S}) as well as the reddening law 
towards NGC\,1893 is found to be average \citep{2007MNRAS.380.1141S}. 
This indicates that size of the dust grains towards NGC\,1893 is comparable to those
in the diffuse ISM.  \citet{2008MNRAS.384.1675J} have also found the presence of average reddening law towards the 
cluster region of Stock\,8 which is also located near the NGC\,1931. 
The indication of slightly bigger dust grains towards NGC\,1931 could be due to relatively 
bigger dust grains within the ``intra-cluster medium". The extinction law in the cluster
region is further discussed in the ensuing section.
 
\subsection{Extinction law}\label{extinctionlaw}

\begin{figure}
\centering
\resizebox{8.275cm}{8.175cm}{\includegraphics{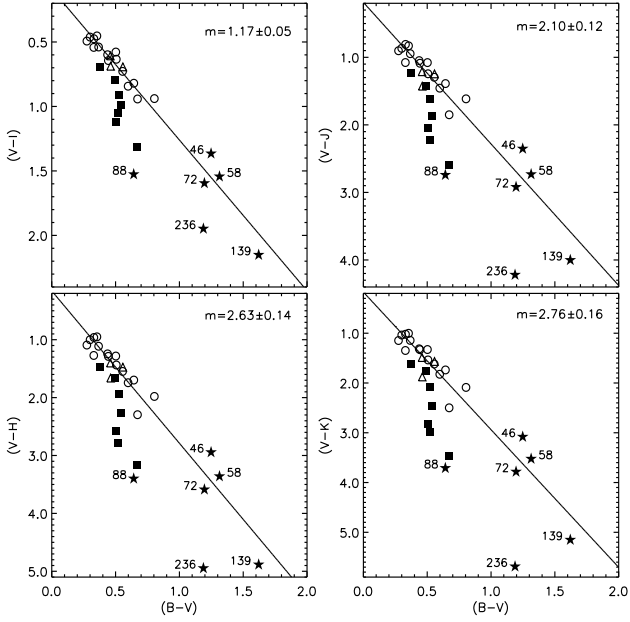}}
\caption{The TCDs of the form $(V-\lambda)$ vs. $(B-V)$, where $\lambda$ is $I$, $J$, $H$ or $K_{s}$. 
Only the probable cluster members identified on the basis of polarimetric data, photometric color-color 
diagrams and color-magnitude diagrams have been used (cf. Sec \ref{membership}). 
The open circles and triangles represent the stars 
having $V(RI)_{C}$ and single or double band polarimetric data, respectively. 
The star symbols are the probable PMS stars (cf. Sec \ref{few_ysos}). 
The filled square symbols are the stars lying in the northern nebulous region. 
The slopes of the straight line fit are indicated on right hand side of each panel. 
The stars in the nebulous region (filled square symbols) and two (\# 88 and \# 236) PMS stars are not used in the fit.}
\label{TCD_polmemb}
\end{figure}

The $(V-\lambda)$ vs. $(B-V)$ TCDs, where $\lambda$ is one of the wavelengths of the broadband 
filters $R, I, J, H, K$ or $L$, have been used to separate the influence of the extinction 
produced by the diffuse ISM from that of the extinction due to the ICM \citep[cf.][]{1990A&A...227..213C,2000PASJ...52..847P}. 
The ($V-\lambda$) vs. ($B-V$) TCDs for the probable members (selected on the basis of polarimetric analysis and optical color-color 
diagrams; cf. Sec \ref{membership}) and probable PMS stars (cf. Sec. \ref{few_ysos}) of the cluster region 
are shown in Fig. \ref{TCD_polmemb}. The open circles are the stars with $V(RI)_{C}$ polarimetric data and the triangles are 
those with either single or double band polarimetric data. 
The slope for the general distribution of majority of the stars (excluding the stars in the nebulous region 
(filled square symbols) and two (\# 88 and \# 236) PMS stars) is found to 
be 1.17$\pm$0.05, 2.10$\pm$0.12, 2.63$\pm$0.14 and 2.76$\pm$0.16
for $(V-I)$, $(V-J)$, $(V-H)$, $(V-K)$ versus $(B-V)$ TCDs respectively. 
These slopes are higher in comparison to those obtained for diffuse ISM, 
which indicates a differing reddening law in the cluster region.
The stars associated with the nebulous region of the northern cluster, shown with filled square symbols, 
seems to be deviate from the distribution of the majority of the stars. 

\begin{figure}
\centering
\resizebox{8.275cm}{9.275cm}{\includegraphics{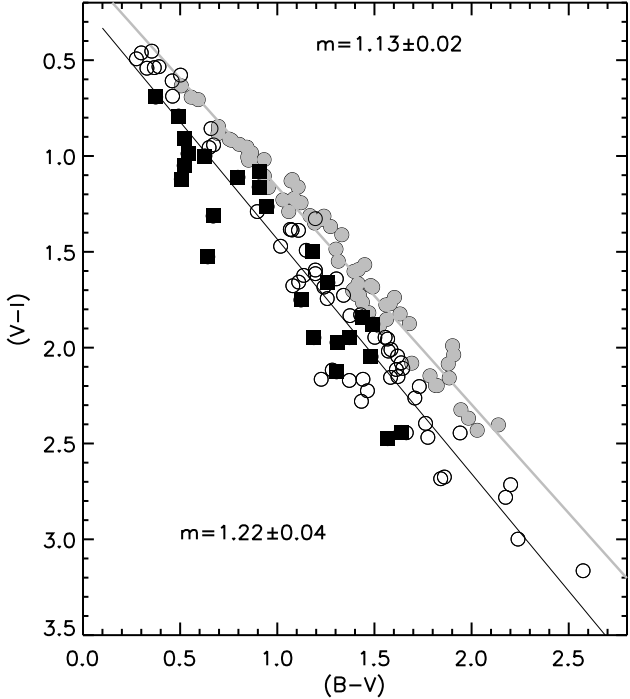}}
\caption{The $(V-I)$ vs. $(B-V)$ TCD for the stars distributed in the cluster region. 
The gray filled circles are considered as field stars, whereas those shown with open circles 
are considered as probable cluster members. The stars distributed in the northern nebulous part of the 
cluster are shown with filled squares. The slopes of the gray and black continuous straight 
line fits to the field (1.13$\pm$0.03) and 
probable cluster members (1.22$\pm$0.04), respectively, are also indicated.}
\label{TCD_photmemb}
\end{figure}

To have a general view of the reddening law in the cluster region ($r$ $\le$ 3.5 arcmin), we used TCDs 
for all the sources detected in the region. Fig. \ref{TCD_photmemb} 
shows $(V-I)$ vs. $(B-V)$ TCD which indicates a combination of 
distributions for the field stars and the cluster members. We selected probable 
field stars visually assuming that stars following the slope of the diffuse ISM are contaminating 
field stars in the cluster region, and they are shown by gray filled circles. 
The slopes for the diffuse ISM have been taken from ~\citet{2003A&A...397..191P}. 
The remaining sources shown by open circles may be probable members of the cluster. 
The probable members associated with the nebulous region are shown by filled squares. 
The slopes of the distributions for probable cluster members (i.e. open circles + filled squares), $m_{cluster}$ are found to be 
1.22$\pm$0.04, 2.20$\pm$0.08, 2.67$\pm$0.10, 2.81$\pm$0.12, 
for the $(V-I)$, $(V-J)$, $(V-H)$, $(V-K)$ vs. $(B-V)$ TCDs respectively. 
The ratios $E(V-\lambda)/E(B-V)$ and the ratio of total-to-selective extinction for the general 
distribution of probable cluster members in the cluster region, $R_{cluster}$, is derived using the 
procedure given by ~\citet{2003A&A...397..191P}. Assuming the value of $R_{V}$ for the diffuse foreground ISM as 3.1,  
the ratios $E(V-\lambda)/E(B-V)$ yield $R_{cluster}$ = 3.3$\pm$0.1, which is in fair agreement 
with the value derived using polarimetric data (cf. Sec \ref{dust_properties}) 
and indicate a differing reddening law in the cluster region. 
The probable members associated with the nebulous region, shown by filled squares, indicate 
a relatively higher $R_V$ value for the nebulous region. Several studies 
e.g. \citet{2003A&A...397..191P}, \citet{2008MNRAS.383.1241P} and references therein, have 
pointed out a high $R_{V}$ values in the vicinity of star-forming regions, 
which is attributed to the presence of larger dust grains in the region.

\subsection{Polarization efficiency}

\begin{figure}
\centering
\resizebox{8.275cm}{8.275cm}{\includegraphics{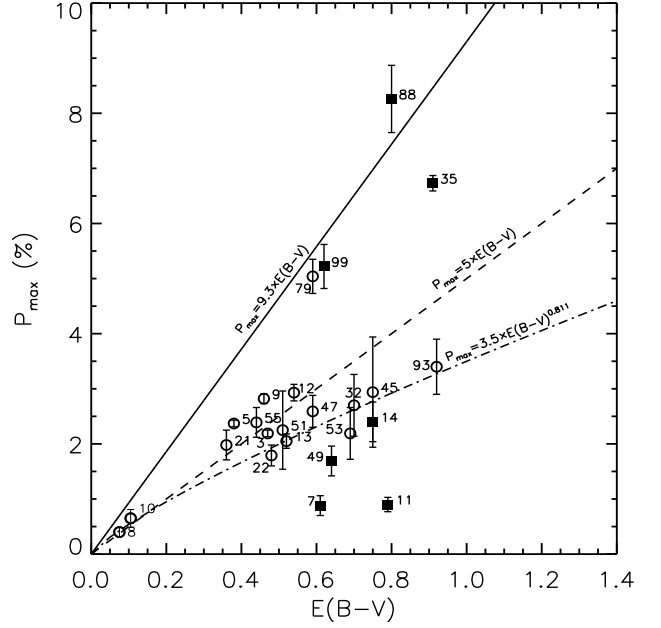}}
\caption{$P_{max}$ versus $E(B-V)$ diagram. The symbols are the same as in Fig. \ref{stokesplane_VRI}. 
The continuous line represents the empirical upper limit relation for the polarization efficiency (assuming $R_V$= 3.1) 
of $P_{max} = 9.3 \times E(B-V)$ \citep{1975ApJ...196..261S}. The dashed line represents the 
relation $P_{max} = 5 \times E(B-V)$ \citep{1975ApJ...196..261S} and the dashed-dotted line represents 
the relation $P_{max}$ = 3.5 $\times$ $E(B-V)^{0.811}$ by \citet{2002ApJ...564..762F}.}
\label{pol_effi}
\end{figure}

The polarization efficiency [$P_{max}/E(B-V)$] of the medium depends mainly on the magnetic field 
strength, the grain alignment efficiency and the amount of depolarization due to the radiation 
passing through various mediums having different magnetic field directions 
\citep[see e.g.][E11 \& E12 and references therein]{2003ApJ...598..349F}. 
Figure \ref{pol_effi} displays the polarization efficiency diagram for the stars towards NGC\,1931. 
The empirical upper limit relation for the polarization efficiency of the diffuse ISM,  given by  
$P_{max}=3A_{V} \simeq 3R_{V} E(B-V) \simeq  9.3 E(B-V)$ ~\citep[assuming $R_{V}$=3.1,][]{1956ApJ...124..367H,1975ApJ...196..261S}, 
shown by a continuous line in Fig. \ref{pol_effi}. \citet{1975ApJ...196..261S} have shown that the polarization efficiency of 
the ISM in general follows a mean relation $P_{max} \simeq 5 E(B-V)$ and the same is shown by a dashed line. 
The dashed-dotted line represents the average polarization efficiency for the general diffuse ISM by \citet{2002ApJ...564..762F}, 
which is valid for $E(B-V)$ $\textless$ 1.0 mag. The majority of the cluster members are distributed below the dashed line 
indicating that polarization efficiency of the ICM is less than the polarization efficiency of the diffuse ISM ($\sim$5 per cent per mag) 
and seems to follow the relation for the general diffuse ISM by \citet{2002ApJ...564..762F}. Four stars, \#35, \#79, \#88 and \#99 
have relatively high polarization ($P_{max}$ $\sim$ 5 to 8 per cent) and polarization efficiency greater than 
5 per cent per mag. Interestingly, the stars \#35 and \#88 have $\sigma_1$ and $\overline\epsilon$ 
values $\textgreater$ 1.5 and $\textgreater$ 2.3 respectively, thereby indicating presence of intrinsic 
polarization and rotation in their polarization angles. It is worthwhile to mention that these two stars 
are associated with the nebulosity of the northern region. The star \#79 has $\overline\epsilon$ 
$\textgreater$ 2.3 which indicates a rotation in the polarization angle. The stars \#7 and \#11 show 
significantly smaller polarization efficiency ($\sim$ 1 per cent per mag). The $\overline\epsilon$ values for these 
stars are $\textgreater$ 2.3, hence there could be a rotation in the polarization angle. 
As we have already mentioned (cf. Sec \ref{distri_p_t}), the polarization measurements of these two stars (\#7 and \#11) 
are less reliable. Interestingly, all the stars deviating from the mean polarization efficiency 
behavior of the region are located in the northern region.

Here it is interesting to mention that the dust grains towards NGC\,1893 ($l=173\fdg6, b=-1\fdg7$), which is spatially 
close to NGC\,1931, show higher polarization efficiency (cf. figure 9, E11)  whereas the 
the dust grains towards NGC\,1931 exhibit less polarization efficiency in comparison to mean polarization 
efficiency for the general diffuse ISM. This could be due to the average extinction law in the foreground medium and 
differing reddening law in the intra-cluster medium.

\section{STELLAR CONTENTS OF THE CLUSTER}\label{stellar_contents}

\subsection{Optical color-magnitude diagram: Distance and Age}\label{dist_age} 

\begin{figure}
\centering
\resizebox{8.275cm}{9.075cm}{\includegraphics{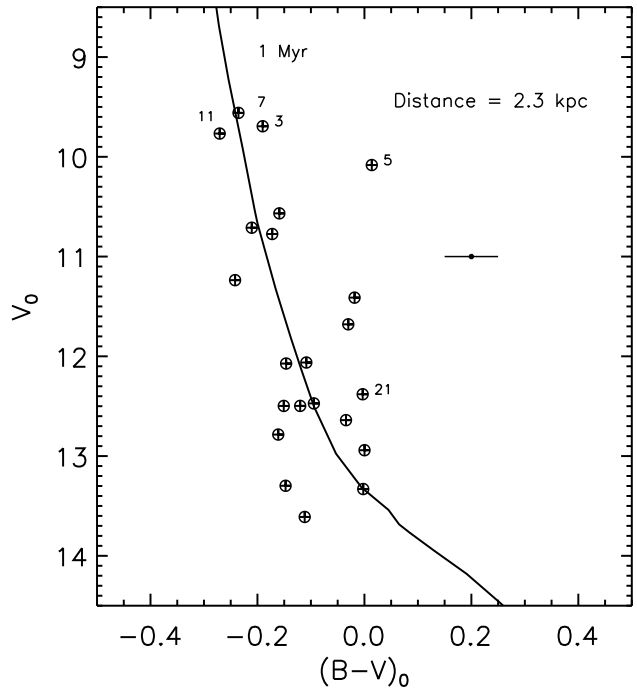}}
\caption{De-reddened $V$ versus $(B-V)$ CMD for probable member stars (cf. Sec. \ref{membership}). 
The post main-sequence isochrones for 1 Myr (Z=0.02) by \citet{2008A&A...482..883M} has been 
plotted after correcting for the distance and reddening. 
The average error in color is shown in the right side of the figure.
The polarization of two sources \#5 and \#21 is comparable to the cluster members, however
the $E(B-V)$ values for these stars are less than the 
minimum reddening ($E(B-V)_{min}$=0.50 mag) for the cluster region (cf. Table \ref{pol_ccd_members}).} 
\label{BV0V0CMD}
\end{figure}

The  $V_{0}/(B-V)_{0}$ color-magnitude diagram (CMD; Fig. \ref{BV0V0CMD}) 
of the identified 22 probable members (Sec \ref{membership}) has been used to estimate the distance to 
the cluster. Reddening of individual probable members having spectral types earlier than $A0$ 
has been estimated using  the reddening free index Q~\citep{1953ApJ...117..313J} which is defined as $Q = (U-B)-0.72(B-V)$. 
The reddening law was assumed to be normal. 
The intrinsic $(B-V)_{0}$ color and color-excess for the MS stars can be obtained 
from the relation $(B-V)_{0}$ = 0.332$\times$Q ~\citep{1966ARA&A...4..193J,1993AJ....106.1906H} and $E(B-V) = (B-V)-(B-V)_{0}$, respectively. 
A visual fit of the isochrone for 1 Myr and Z=0.02 by \citet{2008A&A...482..883M} to the observations yields a distance 
modulus of $V_{0}-M_{V}$=11.81$\pm$0.3 which corresponds to a distance of 2.3$\pm$0.3 kpc.  
The distance estimate is in agreement with that obtained by \citet{1986Ap&SS.120..107P,1994BASI...22..291B, 2009MNRAS.397.1915B}.

Stars \#7 ($E(B-V)$ =0.61) and \#11 ($E(B-V)$ = 0.79) are located in the northern nebulous region and could
be the ionizing source(s) of the region (cf. Sec \ref{SFR_SCENARIO}). The star \#3 is located in the southern 
region and has $E(B-V)$ =0.47. The present photometric data for star \#3 is in agreement with that
reported by \citet{1979A&AS...38..197M}. The $E(B-V)$ value for star \#3 is comparable to the $E(B-V)_{min}$. 
\citet{1979A&AS...38..197M} concluded that either this star is a foreground star or could be a binary
star. Since polarimetric observations suggest that it could be a member, we presume that this star could
be binary star and a member of the cluster.

\begin{figure}
\centering
\resizebox{8.275cm}{5cm}{\includegraphics{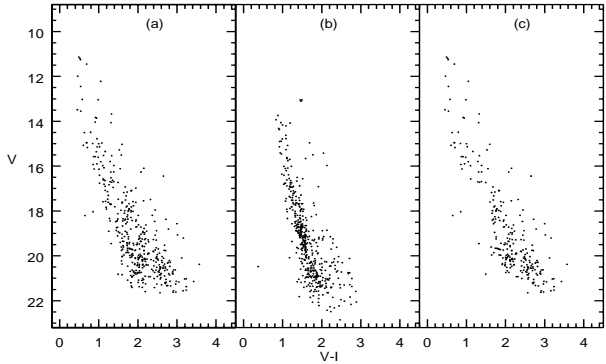}}
\caption{The $V/(V-I)$ CMDs for all the stars within $r_{cl}~\leqslant~3.5\arcmin$
of the cluster region (left panel), the nearby field region (middle panel) and
statistically cleaned $V/(V-I)$ CMD (right panel).}
\label{VVICMD_CLFGCLEANED}
\end{figure}

\begin{figure}
\centering
\resizebox{8.275cm}{8.275cm}{\includegraphics{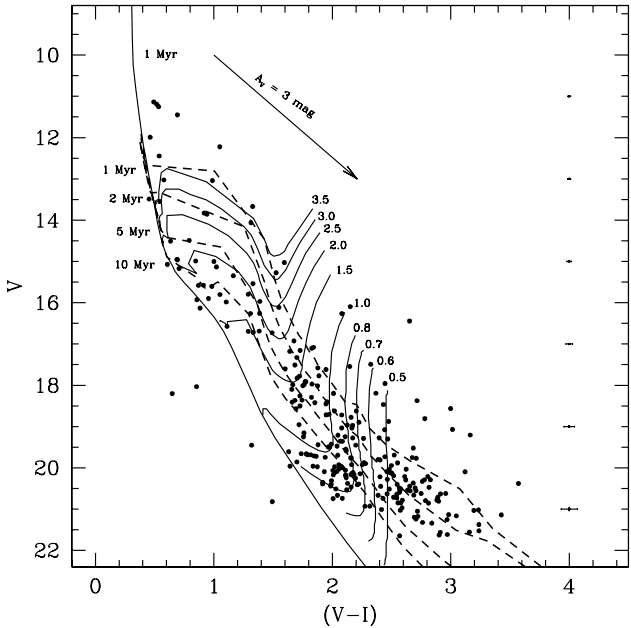}}
\caption{The statistically cleaned
CMD for the region within the 3.5 arcmin radius (cf. Fig. \ref{stellar_density_contours}).
The average error in $V$ and $(V-I)$ colors are shown in the right side of the figure.}
\label{VIVCMD_CLEANED2}
\end{figure}

The $V/(V-I)$ CMDs of all the stars within the cluster region (i.e. $r_{cl}~\leqslant~3.5\arcmin$) and 
the nearby field region are shown in the left and middle panels of Fig. \ref{VVICMD_CLFGCLEANED}. 
As discussed in our earlier works \citep[e.g.,][]{2008MNRAS.383.1241P,2008MNRAS.384.1675J} 
the removal of field star contamination from the sample of stars in the cluster region is necessary as both PMS and dwarf 
foreground stars occupy similar positions above the ZAMS in the CMDs. 
Since proper-motion data is not available for the region, the statistical criterion was used to estimate the 
number of probable member stars in the cluster region. To remove contamination due to field stars, we statistically subtracted 
their contribution from the CMD of the cluster region using the following procedure. The CMDs of the cluster as well 
as of the reference region were divided into grids of $\Delta{V}=1$ mag and $\Delta (V-I)$ = 0.4 mag. 
The number of stars in each grid of the CMDs were counted. After applying the completeness corrections  
using the CF (cf. Table \ref{tab_completeness}) to both the data samples, the probable number of cluster members in each 
grid were estimated by subtracting the corrected reference star counts from the corrected counts in the cluster region. 
The estimated numbers of contaminating field stars were removed from the cluster CMD in the following manner.  
For a randomly selected star in the CMD of the reference region, the nearest star in the cluster CMD 
within $V\pm0.25$ and $(V-I)\pm 0.125$ of the field was removed. Although the statistically 
cleaned $V/(V-I)$ CMD of the cluster region shown in the right panel of Fig. \ref{VVICMD_CLFGCLEANED} clearly 
shows the presence of PMS stars in the cluster, however the contamination due to field stars 
near $V$ $\sim$ 20 mag and $(V-I)$ $\sim$ 2.0 can still be seen (cf. Fig. \ref{VIVCMD_CLEANED2}). 
This field population could be due to the background population as discussed by \citet{2006MNRAS.373..255P}.

The statistically cleaned  $V/(V-I)$ CMD (SCMD) of the cluster region with PMS isochrones by \citet{2000A&A...358..593S}
for ages 1, 2, 5 and 10 Myr (dashed curves) and 1 Myr (continuous curve) by \citet{2008A&A...482..883M} is shown in the
Fig. \ref{VIVCMD_CLEANED2} which manifests the presence of PMS population with an age spread of
$\sim$~5 Myr. The statistics of PMS sources obtained from the SCMD can be used to study the initial mass function ($IMF$) of
PMS population of NGC\,1931. Here we would like to point out that the points shown by filled circles in Fig.  \ref{VIVCMD_CLEANED2}
may not represent the actual members of the clusters. However, the filled circles should represent the
statistics of PMS stars in the region and it has been used to study the $MF$ of the cluster region.

\subsection{Identification of YSOs}\label{identi_ysos} 
\subsubsection{On the basis of $(J-H)/(H-K)$ TCD}\label{nir_ccd}

\begin{figure*}
\centering
\resizebox{8.175cm}{8.175cm}{\includegraphics{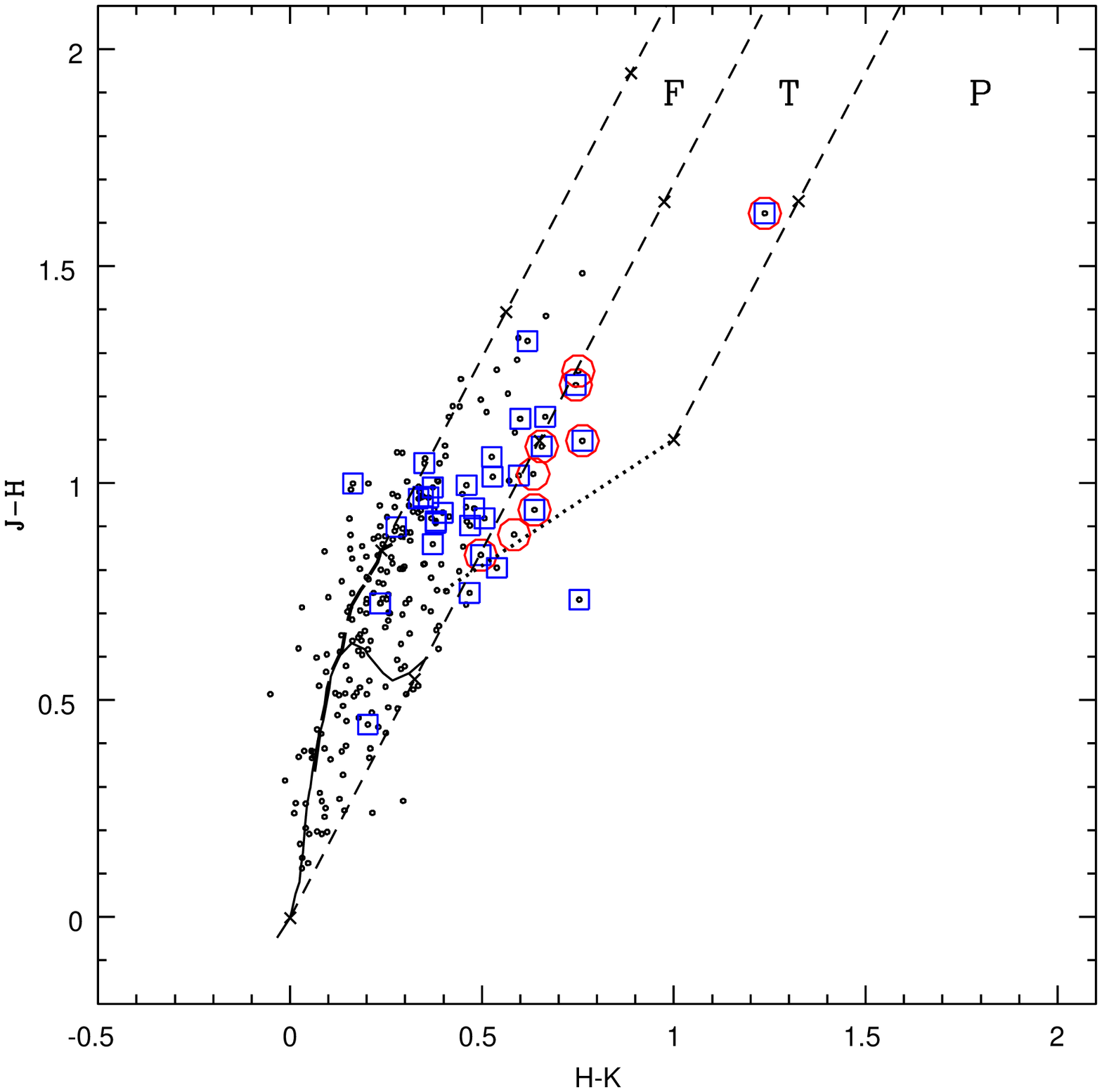}}
\resizebox{8.175cm}{8.175cm}{\includegraphics{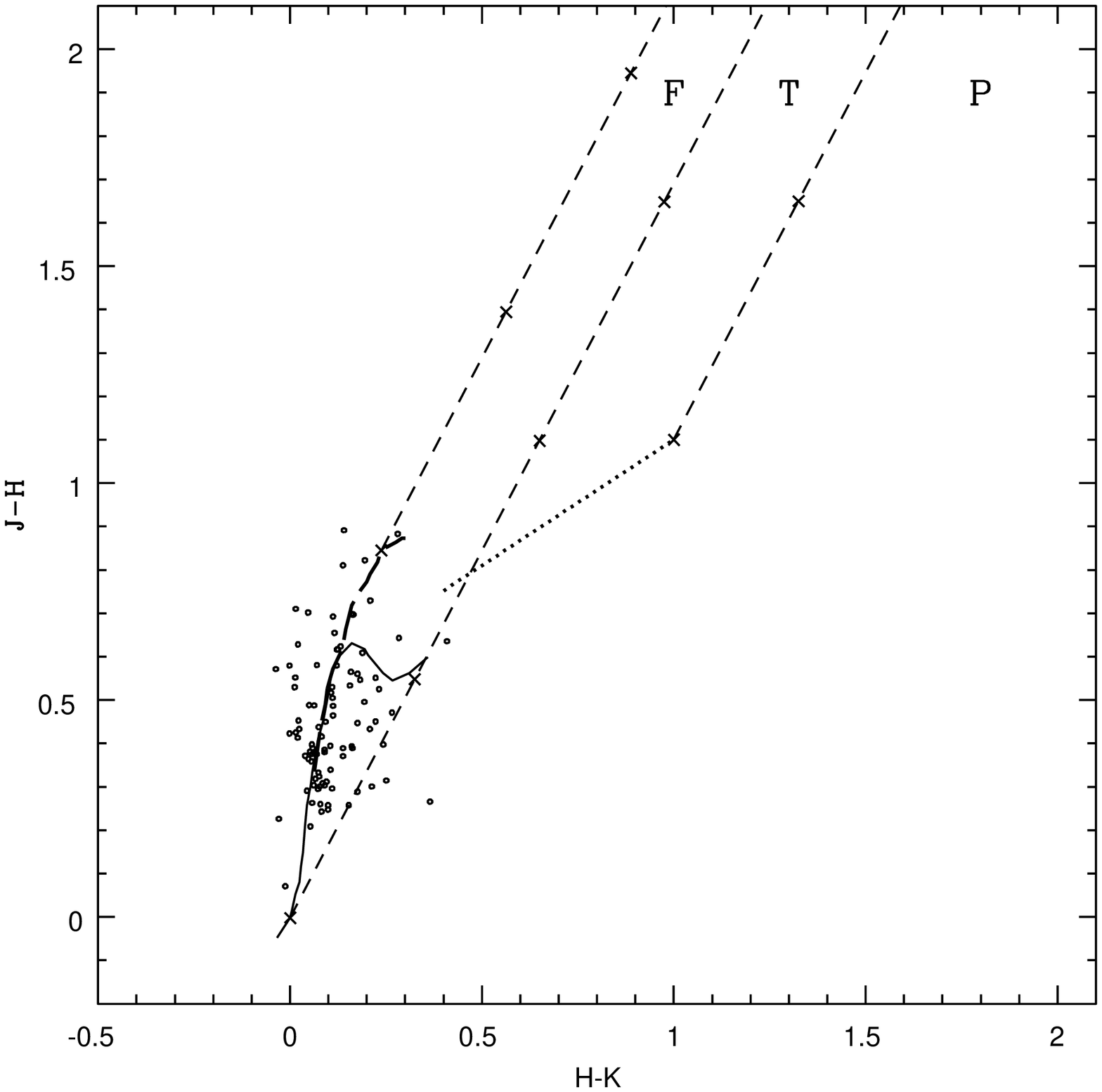}}
\caption{({\it Left panel}): The NIR TCD for the stars in the region of NGC\,1931.
({\it Right panel}): Same as left panel but for the reference region. The continuous and thick
dashed curves represent the unreddened MS and giant branch \citep{1988PASP..100.1134B}, respectively.
The dotted line indicates the locus of unreddened CTTSs \citep{1997AJ....114..288M}.
The parallel dashed lines are the reddening vectors drawn from the tip (spectral type M4)
of the giant branch (left reddening line), from the base (spectral type $A0$) of the MS branch
(middle reddening line) and from the tip of the intrinsic CTTS line (right reddening line).
The crosses on the reddening vectors show an increment of $A_{V}$ = 5 mag.
The sources shown with circles are the NIR excess sources whereas the sources shown with
square symbols are Class II sources identified on the basis of MIR data (cf. Sec \ref{mir_ccd}).}
\label{NIRCMD}
\end{figure*}

NIR imaging surveys are important tools to detect YSOs in star forming regions. The locations of 
YSOs on $(J-H)/(H-K)$ two-color diagram (NIR TCD) are determined to a large extent by their 
evolutionary state. The NIR TCD using the 2MASS data for all the sources lying in the NGC\,1931 region and 
having photometric errors less than 0.1 magnitude is shown in the left panel of Fig. \ref{NIRCMD}. 
All the 2MASS magnitudes and colors have been converted into the California Institute of Technology (CIT) system. 
The solid and thick dashed curves represent the unreddened MS and giant branch \citep{1988PASP..100.1134B}, respectively.  
The dotted line indicates the locus of unreddened TTSs \citep{1997AJ....114..288M}.  All the curves and lines are also 
in the CIT system. The parallel dashed lines are the reddening vectors drawn from the tip (spectral type M4) 
of the giant branch (left reddening line), from the base (spectral type $A0$) of the MS branch 
(middle reddening line) and from the tip of the intrinsic TTS line (right reddening line).  
The extinction ratios $A_J/A_V = 0.265, A_H/A_V = 0.155$ and $A_K/A_V=0.090$ have been taken from \citet{1981ApJ...249..481C}. 
The  sources located in the `F' region (cf. left panel of Fig. \ref{NIRCMD}) could be either field stars 
(MS stars, giants) or Class III and Class II sources with small NIR-excesses, 
whereas the sources distributed in the `T' and `P' regions are considered to be 
mostly Classical TTSs (CTTSs or Class II objects) with relatively large NIR-excesses and 
likely Class I objects, respectively 
\citep[for details see][]{2008MNRAS.383.1241P,2011PASJ...63..795C,2011MNRAS.415.1202C}. 
It is worthwhile to mention also that \citet{2006ApJS..167..256R} have shown that there
is a significant overlap between protostars and CTTSs. The NIR TCD of the NGC\,1931 region 
(left panel of Fig. \ref{NIRCMD}) indicates that a few sources show $(H-K)$ excess and these
are shown by open circles. A comparison of the TCD of the NGC\,1931 region with the NIR TCD of the 
nearby reference region (right panel of Fig. \ref{NIRCMD}) suggests that the sources lying
in the `T' region could be CTTSs/Class II sources.

\subsubsection{On the basis of MIR data}\label{mir_ccd}

\begin{figure}
\centering
\resizebox{8.175cm}{8.175cm}{\includegraphics{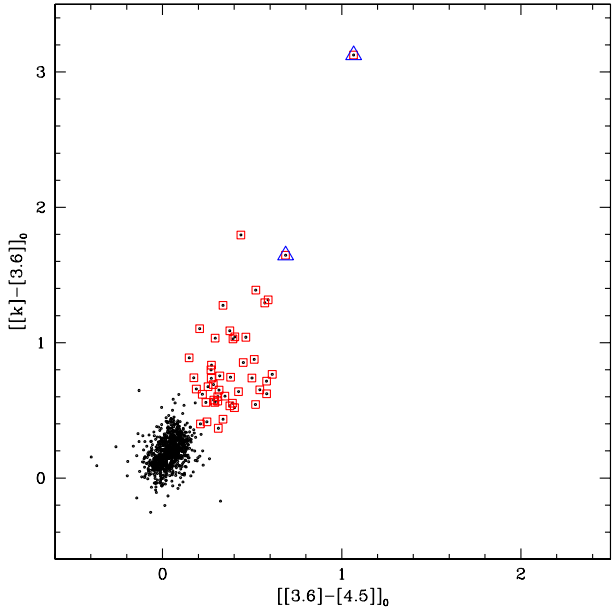}}
\caption{$[[K]-[3.6]]_0$ versus $[[3.6]-[4.5]]_0$ TCD for the IRAC sources having 2MASS K-band counter parts. 
The triangle and squares represent Class I and Class II sources respectively.} 
\label{MIRCCD}
\end{figure}

The {\it Spitzer} MIR observations (3.6 $\mu$m and 4.5 $\mu$m) for the NGC\,1931 region are 
also available. Since young stars in the NGC\,1931 could be deeply embedded, the MIR observations can 
provide a deeper insight into the embedded YSOs.  YSOs occupy distinct regions in the IRAC color plane 
according to their nature; this makes MIR TCDs a very useful tool for the classification of YSOs. Since 
8.0~$\mu$m data are not available for the region, we used $[[K]-[3.6]]_0$ and $[[3.6]-[4.5]]_0$ TCD 
\citep[cf.][]{2009ApJS..184...18G} to identify the deeply embedded YSOs. Fig. \ref{MIRCCD} presents a $[[K]-[3.6]]_0$ versus 
$[[3.6]-[4.5]]_0$ TCD for the observed sources. The identified Class I and Class II sources 
along with photometric data (optical: present work, $JHK_{s}$: 2MASS, 3.6 $\mu$m and 4.5 $\mu$m: {\it Spitzer}, 
3.4 $\mu$m and 4.6 $\mu$m and 12 $\mu$m: WISE) are given in Tables \ref{whole_data_ysos1} \& \ref{whole_data_ysos2}. 
The 3.4 $\mu$m, 4.6 $\mu$m and 12 $\mu$m data 
have been taken from WISE database (http://irsa.ipac.caltech.edu/). 

\subsubsection{On the basis of optical \& NIR CMDs and NIR TCD}\label{few_ysos}

\begin{figure*}
\centering
\resizebox{15cm}{15cm}{\includegraphics{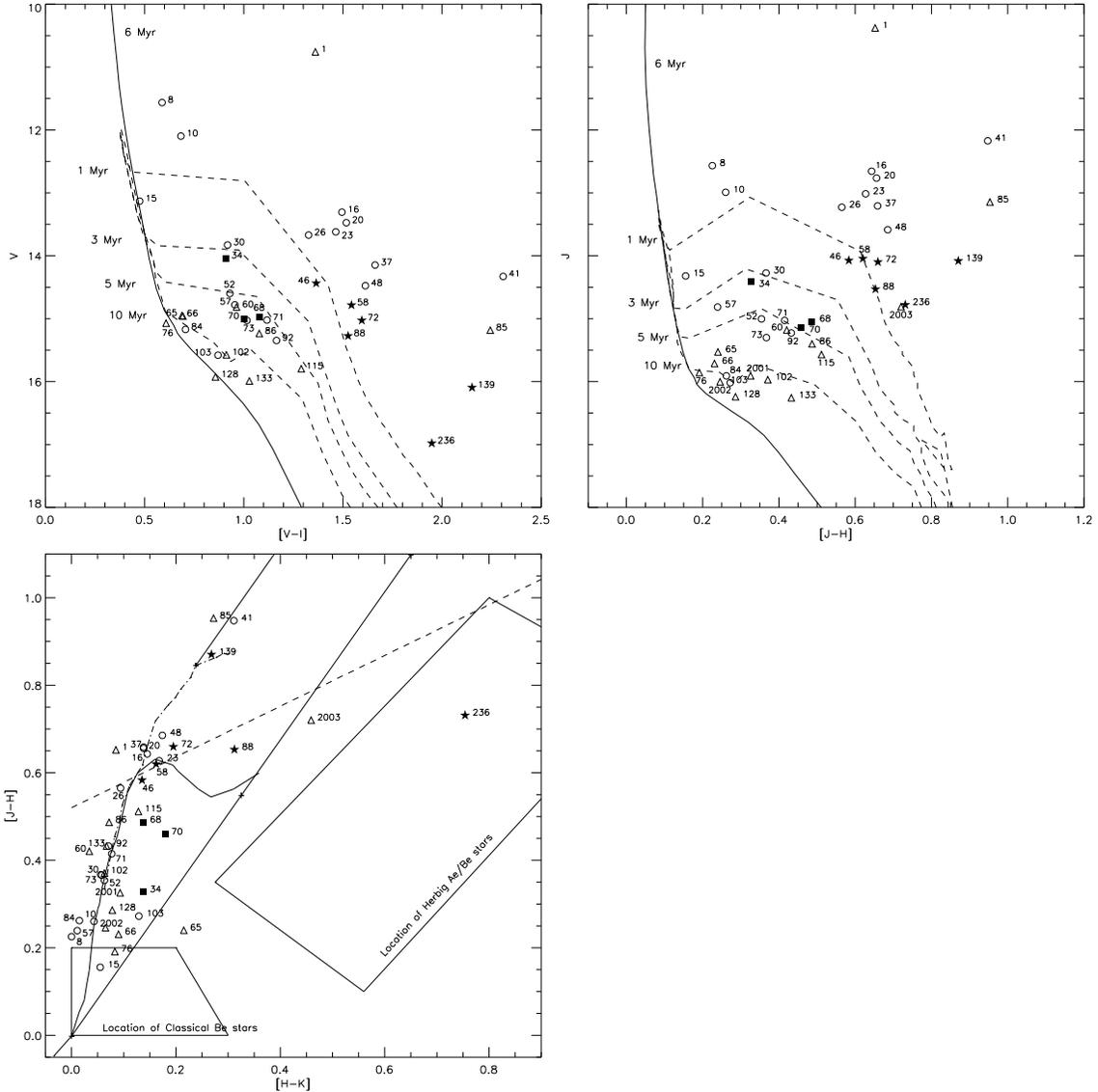}}
\caption{The $V/(V-I)$,  $J/(J-H)$ CMDs and $(J-H)/(H-K)$ TCD for the stars having
polarimetric observations but are not classified as probable members in Sec \ref{membership}.
The stars shown with star symbols are considered as probable members.The symbols are 
same as in Fig. \ref{stokesplane_VRI}.}
\label{CMD_CCD_ALL}
\end{figure*}

Figure \ref{CMD_CCD_ALL} shows $V/(V-I)$ and $J/(J-H)$ CMDs as well as $(J-H)/(H-K)$ two-color-diagram (TCD)
for the remaining stars having polarimetric data but are not classified as probable members in
Sec \ref{membership}. We found a few stars located near 1 Myr isochrone in both the CMDs.
These are shown by star symbols. Four stars \#46, \#58, \#72 and \#88 are
located near the extension of T-Tauri stars (TTSs) locus by Meyer et al. (1997) in $(J-H)/(H-K)$ TCD.
The disk accretion rates $\dot{M}_{disk}$ estimated using the spectral energy distribution
(cf. Sec \ref{accretion_rate}) for the stars \#46 and \#72 are of the order of $10^{-9}$ and $10^{-7}$ $M_{\sun}$/yr
which are comparable to the Class II sources.
We presume these could be weak line TTSs with small NIR excess.
The star \#139 could also be a reddened WTTs, whereas the star \# 236 could be HAe/Be
star. The star \# 236 is identified as a Class II source on the basis of NIR and
MIR data (cf. Sec \ref{mir_ccd}). The $\dot{M}_{disk}$ for \#236 and \#139 is estimated
to be $0.8\times10^{-8}$ $M_{\sun}$/yr and $0.8\times10^{-7}$ $M_{\sun}$/yr, respectively.

\subsection{CMD for the PMS sources}

\begin{figure}
\centering
\resizebox{8.175cm}{8.175cm}{\includegraphics{VI_V_starsin_4arcmin_wt_nebstars_YSO_ctt_clI_clII_North_and_South.epsi}}
\caption{$V$ versus $(V-I)$ CMD for the identified YSOs (Class I, Class II and 
NIR-excess sources respectively shown with open triangles, squares and circles) identified on the 
basis of NIR (cf. Sec \ref{nir_ccd}) and MIR (cf. Sec \ref{mir_ccd}) data. 
The star symbols represent probable YSOs as identified in Sec \ref{few_ysos}. 
The PMS isochrones for 1, 2, 5 and 10 Myr by \citet*{2000A&A...358..593S} 
and isochrone for 1 Myr by \citet{2008A&A...482..883M} are drawn as dashed and continuous curves. All the isochrones 
are corrected for the cluster distance and reddening.
The average error in $V$ and $(V-I)$ colors are shown in the right side of the figure.}
\label{VIVCMD_CLEANED1}
\end{figure}

The $V/(V-I)$ CMD for the YSOs identified in Sec \ref{identi_ysos} 
is shown in Fig. \ref{VIVCMD_CLEANED1}. 
PMS isochrones by Siess et al. (2000) for 1, 2, 5 and 10 Myr (dashed lines) and post-main sequence isochrone 
for 1 Myr by \citet{2008A&A...482..883M} (continuous curve) 
corrected for cluster distance (2.3 kpc) and minimum reddening ($E(B-V)$=0.5) are also shown. 
Figure \ref{VIVCMD_CLEANED1} reveals that a majority of the sources have ages $\textless$ 5 Myr with 
a possible age spread of $\sim$5 Myr. Since the reddening vector in $V/(V-I)$ CMD is nearly parallel to the PMS isochrone, 
the presence of variable extinction in the region will not affect the age estimation significantly. 
The age and mass of each YSO have been estimated using the V/$(V-I)$ CMD, as discussed by
~\citet{2008MNRAS.383.1241P} and ~\citet{2009MNRAS.396..964C} 
and are given in Table \ref{age_and_mass_neelam}. It is worthwhile to point out
that the estimation of the ages and masses of the PMS stars by comparing their locations in the CMDs
with the theoretical isochrones is prone to random as well as systematic errors 
~\citep[see e.g.][]{2005astro.ph.11083H,2008ASPC..384..200H,2009MNRAS.396..964C,2011MNRAS.415.1202C}. 
The effect of random errors due to photometric errors and reddening estimation in determination 
of ages and masses has been estimated by propagating the random errors to their observed estimation 
by assuming normal error distribution and using the Monte-Carlo simulations 
\citep[cf.][]{2009MNRAS.396..964C}. The systematic errors could be due to the use of different PMS evolutionary models
and the error in distance estimation etc. Since we are using evolutionary models by \citet{2000A&A...358..593S} 
to estimate the age of all the YSOs in the region, we presume that the age estimation is affected only by
the random errors. The presence of binaries may also introduce errors in the age determination.
Binarity will brighten the star, consequently the CMD will yield a lower age estimate. In the case of an
equal mass binary we expect an error of $\sim50-60 \%$ in the PMS age estimation. However, it is difficult
to estimate the influence of binaries/variables on mean age estimation as the fraction of 
binaries/variables is not known. In the study of TTSs in the \hii region IC\,1396, \citet{2011MNRAS.415..103B} pointed 
out that the number of binaries in their sample of TTSs could be very low as close binaries loose their
disk significantly faster than single stars \citep[cf.][]{2006ApJ...653L..57B}. Estimated ages and masses of the
YSOs range from $\sim$ 0.1 to 5 Myr and $\sim 0.3-3.5 M_{\sun}$ respectively, which are comparable with the
lifetime and masses of TTSs. The age spread indicates a non-coeval star formation in this region.

\subsection{Physical state of the YSOs} \label{accretion_rate}

\begin{figure}
\centering
\resizebox{8.175cm}{6.175cm}{\includegraphics{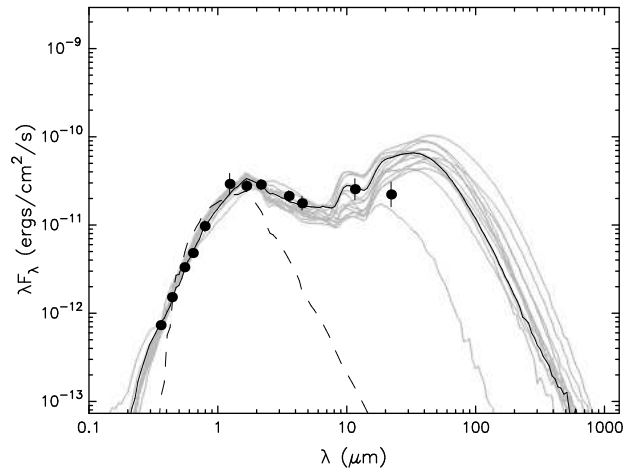}}
\caption{The SED for a Class II source \#236. The black line shows the best fit, and the gray lines show subsequent good fits 
with $\chi^{2}$-$\chi^{2}_{min} \textless$ $2N_{data}$. The dashed line shows the stellar photosphere corresponding 
to the central source of the best fitting model. The filled circles denote the input flux values using the data given 
in Tables \ref{whole_data_ysos1} \& \ref{whole_data_ysos2}.}
\label{SED_236}
\end{figure}

To characterize the circumstellar disk properties of YSOs, we analyzed
the spectral energy distributions (SEDs) using the fitting 
tool of \citet{2007ApJS..169..328R}. The SED fitting tool fits thousands of models to the observed SED
simultaneously. This SED fitting tool determines their physical
parameters like mass (M$_{\star}$), age, interstellar extinction ($A_{V}$), 
temperature ($T_{\star}$), disk mass, disk mass accretion rate ($\dot{M}_{acc}$), etc. The SED fitting tool fits each
of the models to the data, allowing both the distance and foreground extinction to be free
parameters. Since we do not have spectral type information for the identified YSOs, we estimated $A_V$ by 
tracing back sources along the reddening vector to the intrinsic locus of TTS \citep{1997AJ....114..288M} 
in $(J-H)/(H-K)$ TCD. Thus estimated value of $A_{V}$ is considered as the maximum $A_{V}$ whereas the 
foreground extinction for the YSOs, towards NGC\,1931 is considered as the minimum $A_{V}$. 
The distance range is adopted as 2.1 to 2.5 kpc. The error in NIR and MIR flux estimates due to possible uncertainties in the
calibration, extinction, and intrinsic object variability was set as 10$-$20\%.
Figure \ref{SED_236} shows an example of SED of the resulting model for a Class II source \#236. We obtained physical
parameters for all the sources adopting an approach similar to that of \citet{2007ApJS..169..328R} by
considering those models that satisfy $\chi^{2}$-$\chi^{2}_{min} \textless$ $2N_{data}$, where $\chi^{2}_{min}$ is the
goodness-of-fit parameter for the best-fit model and $N_{data}$ is the number of observational input data
points. The relevant parameters are obtained from the weighted mean and the standard deviation of
these models, weighted by e$^{({{-\chi}^2}/2)}$ of each model and are shown in Table \ref{age_and_mass_manash}, which reveals that
majority of the Class II sources have disk accretion rate of the order of $\sim$ $10^{-7}-10^{-8}$ $M_{\sun}$ yr$^{-1}$.
The parameters support the classification of YSOs obtained on
the basis of CMDs and NIR/MIR TCDs. We would like to mention that the stellar ages given in Table \ref{age_and_mass_manash} are only
approximate and in the absence of far-infrared (FIR) to millimeter data the disk parameters,
should be taken with a caution, however these parameters can still be used as a quantitative indicator of stellar youth.

\section {INITIAL MASS FUNCTION AND K-BAND LUMINOSITY FUNCTION} \label{IMF_KLF}

The distribution of stellar masses for a star formation event in a given volume of space
is known as initial mass function (IMF) and in combination with star formation rate, the
IMF dictates the evolution and fate of galaxies and star clusters \citep{2002Sci...295...82K}. The MF of young
clusters can be considered as the IMF as they are too young to lose significant number of members either by 
dynamical or by stellar evolution.

The MF is often expressed by a power law, $N (\log m) \propto m^{\Gamma}$ and 
the slope of the MF is given as: 
$ \Gamma = d \log N (\log m)/d \log m  $
\noindent
where $N  (\log m)$ is the number of stars per unit logarithmic mass interval. For the mass range 
0.4~$\textless$~$M/M_\odot$~$\leqslant 10$, the classical value derived by \citet{1955ApJ...121..161S} 
for the slope of MF is $\Gamma = -1.35$. 

\begin{figure}
\centering
\resizebox{8.275cm}{11.275cm}{\includegraphics{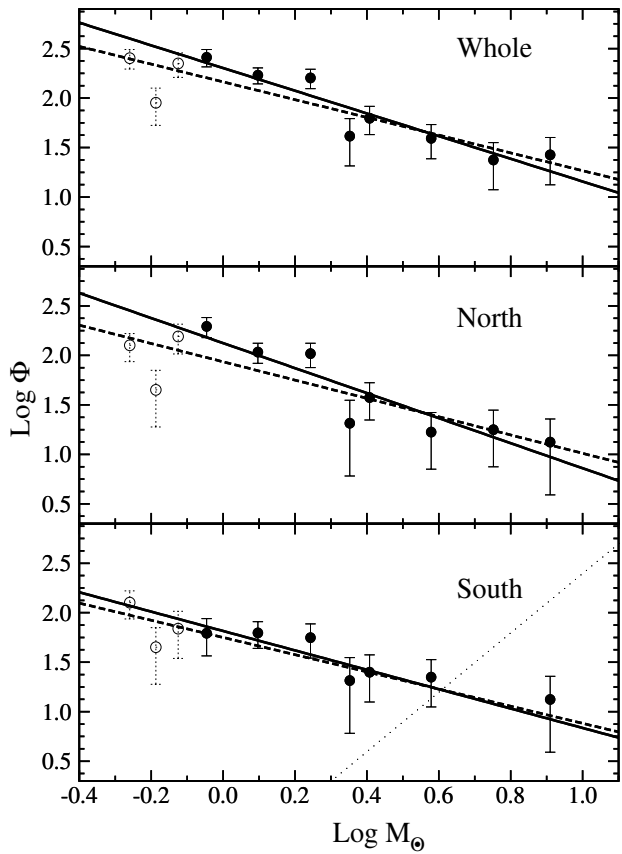}}
\caption{The MF for the whole cluster region (upper panel), the northern (middle panel) and 
southern (bottom panel) regions. The continuous line is the fit to the data excluding the points 
shown by open circles, whereas the dashed line shows fit for the entire sample.}
\label{MF_WHOLENORTHSOUTH}
\end{figure}

With the help of the statistically cleaned CMD, shown in Fig. \ref{VIVCMD_CLEANED2}, 
we can derive the MF using the theoretical evolutionary models. Since the age of the massive cluster 
members could be $\sim$ 5 Myr, the stars having V~$\leqslant$~15 mag ($V_0$~$\leqslant$~13.5 mag; M~$\geqslant$~$2M_\odot$) 
are considered as MS stars. For these stars, the LF was converted to a 
MF using the theoretical models by \citet{2008A&A...482..883M} 
\citep[cf.][]{2001A&A...374..504P,2005MNRAS.358.1290P}. The MF for the PMS stars have been obtained 
by counting the number of stars having age $\leqslant$~5 Myr in various mass bins shown as evolutionary tracks 
in Fig. \ref{VIVCMD_CLEANED2}. The resulting MF for the whole cluster region, 
the northern and southern regions (cf. Fig. \ref{stellar_density_contours}) is plotted in Fig. \ref{MF_WHOLENORTHSOUTH}.
The slope ($\Gamma$) of the MF in the mass range 0.5~$\textless$~$M/M_\odot$~$\textless$~9.5 
for the northern, southern clusters and for the combined region comes out to be $-0.92\pm0.21$, $-0.87\pm0.15$ 
and $-0.90\pm0.16$ respectively, which seems to be shallower than the \citet{1955ApJ...121..161S} value (-1.35). 
However, a careful look of the MF in the northern cluster reveals a break in the power law at $\sim$~0.8$-$1.0~$M_\odot$. 
Excluding the points shown by open cirlces, the slope of the MF in the range 
0.8~$\textless$~$M/M_\odot$~$\textless$ 9.8 is found to be -1.26$\pm$0.23, -0.98$\pm$0.22  
and -1.15$\pm$0.19 for the northern, southern and for the whole cluster regions, respectively. 
The break in the MF slope has also been reported in the case of two clusters, namely 
Stock\,8 and NGC\,1893 located near the NGC\,1931. In the case of young clusters, Stock\,8 \citep{2008MNRAS.384.1675J} 
and NGC\,1893 \citep{2007MNRAS.380.1141S}, the break was reported at $\sim$$1-2$$M_\odot$.  

\begin{figure}
\centering
\resizebox{8.175cm}{6.875cm}{\includegraphics{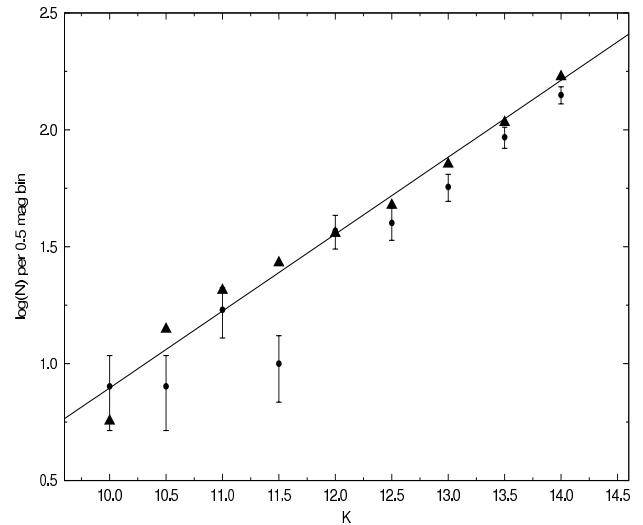}}
\caption{Comparison of the observed KLF (filled circles) and the simulated KLF (triangles) from the 
Galactic model (see text). The error bars represent $\pm\sqrt{N}$ errors.}
\label{KLF_field_model}
\end{figure}

\begin{figure}
\centering
\resizebox{8.275cm}{11.275cm}{\includegraphics{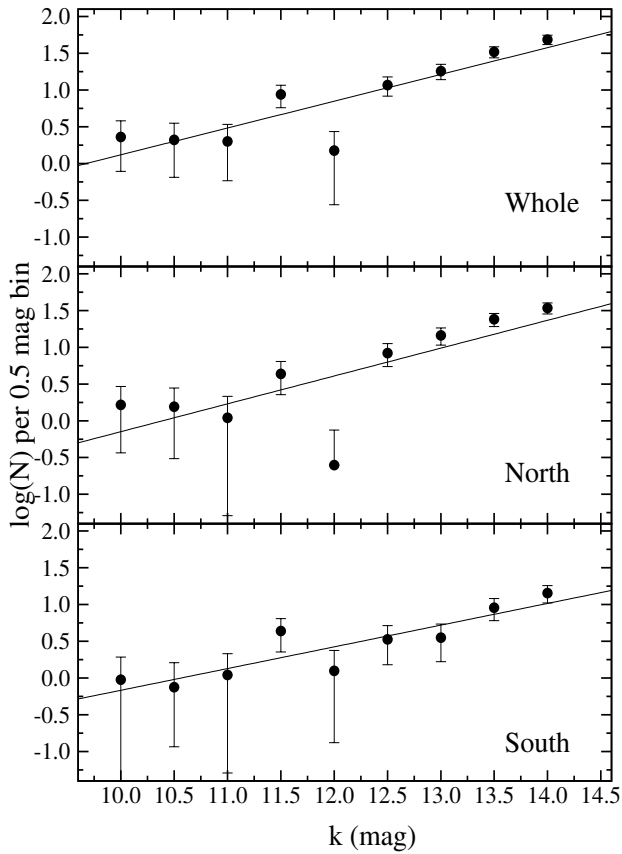}}
\caption{The corrected KLF for North, South and whole cluster regions. The straight line is the least-square fit 
to the data. The error bars represent $\pm\sqrt{N}$ errors.}
\label{KLF_WHOLENORTHSOUTH}
\end{figure}

During the last decade several studies have focused on determination of the $K$-band luminosity 
function (KLF) of young open clusters \citep[e.g][]{2003ARA&A..41...57L,2004ApJ...608..797O,2007MNRAS.380.1141S,2007ApJ...667..963S,2008MNRAS.383.1241P,2008MNRAS.384.1675J,2011MNRAS.411.2530J}. The KLF is being used to 
investigate the IMF and star formation of young embedded clusters. In order to obtain the KLF 
in the region, we have assumed that the NIR data is 99\% complete up to $\sim$ 15.7, 15.1 and 14.3 mag in 
$J$, $H$ and $K_{s}$ bands, respectively as mentioned in Sec \ref{2mass_data}. The Besan\c{c}on Galactic model of stellar population synthesis 
\citep{2003A&A...409..523R} was used to estimate the foreground/background field star contamination 
in the present sample. The star counts both towards the cluster region and towards the direction of the reference field 
we estimated and checked the validity of the simulated model by comparing the model KLF with that of the reference 
field and found that the two KLFs match rather well (Fig. \ref{KLF_field_model}).  
The advantage of this method is that we can separate the foreground ($d~\textless~2.3$ kpc) 
and the background ($d~\textgreater~2.3$ kpc) field star contamination. The foreground 
extinction towards the cluster region is found to be $A_{V} \sim $ 1.6 mag. The model simulations with $d~\textless~2.3$ kpc 
and $A_V$ = 1.6 mag give the foreground contamination, and that with $d~\textgreater~2.3$ kpc and $A_V$ = 2.8 mag give the 
background population. We thus determined the fraction of the contaminating stars (foreground+background) 
over the total model counts. This fraction was used to scale the nearby reference region and subsequently 
the modified star counts of the reference region were subtracted from the KLF of the cluster to obtain the 
final corrected KLF.
This KLF is expressed by the following power-law:

${{ \rm {d} N(K) } \over {\rm{d} K }} \propto 10^{\alpha K}$

\noindent
where ${ \rm {d} N(K) } \over {\rm{d} K }$ is the number of stars per 0.5 magnitude bin 
and $\alpha$ is the slope of the power law. Fig. \ref{KLF_WHOLENORTHSOUTH} shows the KLF for the cluster region 
which yields a slope of $\alpha$ = 0.38$\pm$0.13, 0.29$\pm$0.06 and 0.36$\pm$0.08 for the northern, 
southern regions and for the whole cluster region, respectively. The slopes in the northern and southern regions within the 
error are rather same and similar to the average slopes ($\alpha \sim 0.4$) for young clusters 
~\citep{1991ASPC...13....3L,1995AJ....109.1682L,2003ARA&A..41...57L} 
but higher than the values (0.27$-$0.31) obtained for Be\,59 \citep{2008MNRAS.383.1241P}, 
Stock\,8 \citep{2008MNRAS.384.1675J} and NGC\,2175 region \citep{2012MNRAS.424.2486J}.

\section {Star formation scenario in the NGC\,1931 region} \label{SFR_SCENARIO}

The massive stars in star-forming regions have strong influence and can significantly affect the entire 
region. The star formation in the region may be terminated as energetic stellar winds from massive stars can evaporate nearby clouds.  
Alternatively, stellar winds and shock waves from a supernova explosion may squeeze molecular clouds and induce next generation of star formation. 
\citet{1977ApJ...214..725E} propose that the expanding ionization fronts from the massive star(s) 
play a constructive role in inciting a sequence of star formation activities in the neighborhood. 
The distribution of YSOs and the morphological details of the environment around NGC\,1931 can be used 
to probe the star formation scenario in the region. 

\begin{figure*}
\centering
\resizebox{15cm}{13cm}{\includegraphics{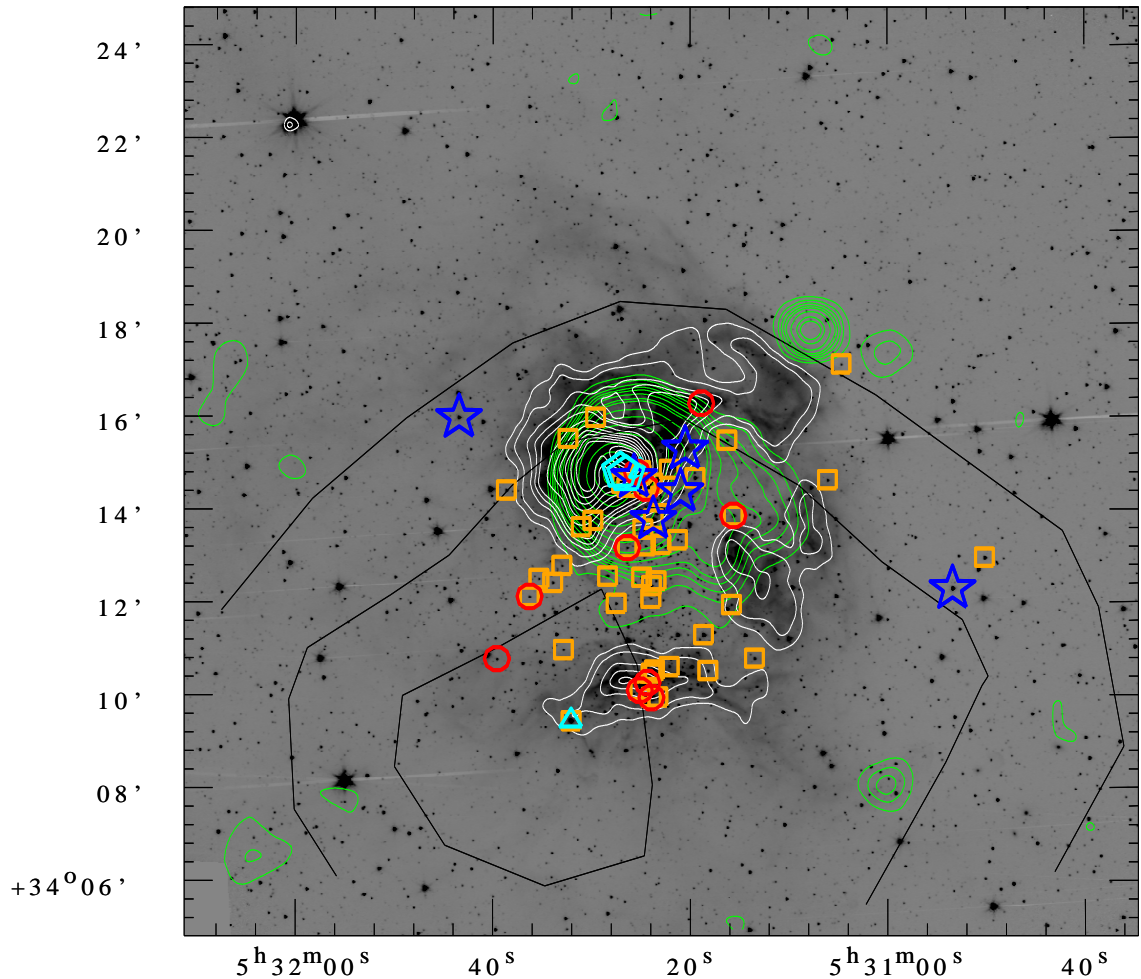}}
\caption{Spatial distribution of Class II sources (square symbols), Class I sources (triangles), 
NIR excess sources (circles) and probable PMS sources (open star symbols) overlaid on K-band image. 
The open pentagon symbols denote two ionizing stars \#7 and \#11. 
The black continuous contours represent $^{12}$CO emission from \citet{1989ApJS...70..731L}. The NVSS 
1.4 GHz radio emission contours (thin green contours) and MSX A-band intensity contours (thick white contours) 
have also been shown. The minimum and maximum levels of NVSS contours are 0.001 Jy/beam and 0.187 Jy/beam 
respectively, whereas the minimum and maximum levels of MSX A-band contours are 2.01$\times$10$^{-6}$ w/m${^2}$-Sr 
and 2.64$\times$10$^{-5}$ w/m${^2}$-Sr respectively. 
North is up and east is to the left. Abscissa and ordinate show the RA (J2000) and Dec (J2000), respectively.
Color version of the figure is available in online journal.}
\label{fig_nvss}
\end{figure*}

In Fig. \ref{fig_nvss}, the $NRAO$ VLA Sky Survey (NVSS) 1.4 GHz radio continuum emission contours 
(thin green contours) and MSX $A$-band (8.3 $\mu$m) MIR contours (thick white contours) along with distribution of YSOs 
are overlaid on the IRAC 3.4 $\mu$m image. In literature we could not find any detailed observations to probe the molecular 
cloud associated with NGC\,1931. The only available is the $^{12}$CO contour map of Sh2-237 region traced 
from figure 27(c) of \citet*{1989ApJS...70..731L}, which is also shown in Fig. \ref{fig_nvss} 
as black contours. Although the resolution is poor, a moderate-sized molecular cloud 
\citep*[NGC\,1931C;][]{1989ApJS...70..731L} of $\sim 6.9\times 10^3 M_\odot$ is 
found to be associated with the cluster. The peak of the $^{12}$CO contours is located at the south-east 
of the cluster center, with a peak value of $H_{2}$ column density $N(H_{2})$ of $\sim 3 \times 10^{21} cm^{-2}$.

The morphology of the radio continuum emission contours reveals that the eastern region is an 
ionization-bounded zone, whereas the decreasing intensity distribution towards the south-west 
direction suggests that this region could be density bounded. There seems to be 
a flow of ionized gas in the south-west direction, as expected in case of Champagne flow model 
\citep{1979A&A....71...59T}. 

Sh2-237 is rather a spherical \hii region around ionizing star(s) and is nearly surrounded by a
dust ring, as revealed by MIR emission in the MSX A-band 
(centered at 8.3 $\mu$m) as well as in the {\it Spitzer IRAC} channel 2 (centered at 4.5 $\mu$m).
The far-ultraviolet (UV) radiation can escape from the \hii region and penetrate the surface of molecular clouds
leading to the formation of photo-dissociation
region (PDR) in the surface layer of the clouds. Polycyclic aromatic hydrocarbons (PAHs) within the PDR are
excited by the UV photons re-emitting their energy at MIR wavelengths, particularly between
6 and 10 $\mu$m. The A-band (8.3 $\mu$m) of MSX includes several discrete PAH emission features
(e.g., 6.6, 7.7, and 8.6 $\mu$m) in addition to the contribution from the thermal continuum component
from hot dust. The ring of PAH emission lies beyond the ionization front (IF), indicating the interface between
the ionized and molecular gas (i.e. PDR). The absence of 8 $\mu$m emission in the interior of the
\hii region is interpreted as the destruction of PAH molecules by intense UV radiation of the ionizing star.

The 1.4 GHz radio continuum emission contours derived from the NVSS \citep{1998AJ....115.1693C} reveal a peak 
around $\alpha_ {2000} \sim 05^{h} 31^{m} 28^{s}$ and $\delta_{2000} \sim +34\degr 14\arcmin 48\arcsec$. 
The integrated flux density of the radio continuum 
above $3\sigma$ level is estimated to be 0.76 Jy. 
\citet{1989ApJS...70..731L} have reported the flux density of the \hii region as 0.6 Jy. 
We estimated the number of Lyman continuum photons ($N_{Lyc}$) 
emitted per second, and hence the spectral type of the exciting star using the flux 
density of 0.76 Jy together with the assumed distance and by assuming a spherical symmetry for the ionized region 
and neglecting absorption of UV radiation by dust inside the \hii region. Assuming a typical electron temperature 
in region as  8000 K \citep[see e.g.,][]{2007ApJ...671..555S}, the relation given by \citet*{2003A&A...407..957M} yields 
log $N_{Lyc}$ = 47.26, which corresponds to a MS spectral type of $\sim$B0.5 \citep{1973AJ.....78..929P,1997A&A...322..598S}. 
The region has two brightest stars (cf. Fig. \ref{BV0V0CMD};  \#7,  $M_V$ $\sim$ -2.24 and \#11,  $M_V$ $\sim$ -2.03; 
spectral type $\sim$ B2) associated with the northern nebulosity. 
If these stars are the ionization sources of the region, the $N_{Lyc}$ produced by these stars will be 
$log(N_{Lyc}) \approx 45.0$. The $N_{Lyc}$ derived from 1.4 GHz radio observations are quite high 
relative to the available stellar content estimated from the broad-band photometry.

The spectral type of the two brightest stars (probable ionizing stars) is found to be $\sim$ B2V. 
The stars seem to be on the MS. The MS life-time of B2 type star 
is $\sim$25 Myr \citep{2006S&W....45i..96L}. This suggests that the \hii region may still be under expansion. 
The dynamical age of Sh2-237 can be estimated using the model by \citet{1997pism.book.....D} with an 
assumption that the ionized gas of pure hydrogen is at a constant temperature in an uniform medium of constant density. 
If the \hii region associated with Sh2-237 has $N_{Lyc}$ $\sim1.855 \times 10^{47}$ (log$N_{Lyc}$ $\sim$~47.26) 
effective ionizing photons per second and is expanding 
into a homogeneous medium of typical density $10^4 (10^3)~ {cm}^{-3}$, 
it will form a Str$\ddot{o}$mgren radius ($r_{s}$) of $\sim$ 
0.022(0.10)~pc. The initial phase is overpressured compared to the neutral gas and will expand into 
the medium of radius $r_{i}$ at time $t$ as 
\[
r_{i}=r_{s} \left( 1+ \frac{7~C_{II}~t}{4~r_{s}} \right)^{4/7}
\]

The observed radius of the ionized gas from NVSS radio contours is found to be $\sim 2\farcm4$ (cf. Fig. \ref{fig_nvss}), 
which corresponds to a physical radius of about 1.6 pc at a distance of 2.3 kpc. In the 
literature the expansion velocity of the ionized front is reported as $\sim$ 4 km s$^{-1}$ 
~\citep[e.g.,][]{1993A&A...273..619M,2009ApJ...701..454K,2009A&A...494..987P}. 
Assuming an expansion speed ($C_{II}$) of 4 km~s$^{-1}$ and typical density of the ionized gas as $10^4 (10^3)~ cm^{-3}$, 
the dynamical age of the ionized region comes out to be $\sim$ 6 (2) Myr. The estimated dynamical age must 
be considered highly uncertain because of the assumption of an uniform medium, assumed velocity of 
expansion etc. Since the \hii region is found to be partially surrounded by the MSX dust emission, 
we would expect an expanding IF to be preceded by a swept-up shell of cool neutral gas as it erodes 
into a neutral cloud.

Fig. \ref{fig_nvss} also shows the distribution of identified YSOs. 
The distribution of YSOs does not seem to follow the stellar density distribution 
of the cluster region and hence we presume that these YSOs could be next generation stars. 
A significant number of YSOs are found to be associated with the dust ring as revealed by MIR emission 
in the MSX A-band (cf. Fig. \ref{fig_nvss}). 
This could be an indication of triggered star formation. In the literature two mechanisms 
which could be responsible for triggered star formation have been discussed. These are 
``radiation-driven implosion" (RDI) and ``collect-and-collapse" models. 

Detailed model calculations of the RDI process have been carried out by several authors 
~\citep[e.g.,][]{1989ApJ...346..735B,1995A&A...301..522L,1997A&A...324..249L,2002ApJ...577..798D,2003MNRAS.338..545K,2006MNRAS.369..143M}. 
In this process, a pre-existing dense clump is exposed to the ionizing radiation from massive stars 
of the previous generation. The head part of the clump collapses due to  the high pressure of the 
ionized gas and the self-gravity, which consequently leads to the formation of next generation stars. 
The signature of the $RDI$ process is the anisotropic density distribution in a relatively small 
molecular cloud surrounded by a curved ionization/shock front (bright rim) as well as a 
group of YSOs aligned from the bright rim to the direction of the ionizing source. 

In the ``collect and collapse" process, which was proposed by 
~\citet{1977ApJ...214..725E}, as the \hii region expands, the surrounding neutral material is 
collected between the ionization front and the shock front which precedes the former. 
With time the layer gets massive and consequently becomes gravitationally unstable and 
collapses to form stars of the second generation including massive stars. Recent simulations 
of this process are shown by \citet{2005ApJ...623..917H,2006ApJ...646..240H} and \citet{2007MNRAS.375.1291D}. 
An observational signature of the process is the presence of a dense layer and massive condensations adjacent to  
an \hii region \citep[e.g.,][]{2003A&A...399.1135D}.

The YSOs do not show any aligned distribution as expected in the case of RDI and hence the 
RDI cannot be the process for triggering of the next generation of stars. 
Since the probable ionization sources in the region have spectral types of $\sim$B2, the maximum MS life time 
of a B2 type star could be $\sim$25 Myr \citep[cf.][]{2006S&W....45i..96L}. The dynamical age of the ionized region 
is estimated in the range of $\sim$2-6~Myr. The spherical morphology of the ionized region 
partially surrounded by MSX dust emission in the mid-IR (see Fig. \ref{fig_nvss}) and association of YSOs 
(having mean age $\sim$ 2$\pm$1 Myr) with the dust region indicate triggered star formation which could be due to 
``collect-and-collapse" process. Some of the YSOs are not found to be associated with the dust and 
are distributed in the central part. 
We presume that these sources may be located in the collected material lying in the outer skin of the 
molecular cloud towards the line of sight of the observer. The $^{12}$CO map by \citet{1989ApJS...70..731L} 
also suggests that the cluster is located at the extreme end of the molecular cloud. The overall star formation 
scenario in the NGC\,1931 region also seems to indicate for a blister model of star formation as suggested by 
\citet{1973ApJ...183..863Z}. \citet{1978A&A....70..769I} has also found that the majority of \hii regions were located at 
the edges of molecular clouds and have structure suggestive of IF surrounded by less dense envelopes.

However, in a recent model calculation, \citet{2011arXiv1109.3478W} argued that the clumps around the
edge of \hii regions do not require the collect and collapse process and can be explained by the
pre-existing non-uniform density distribution of the molecular cloud into which the \hii region expands.
This star formation scenario can also be speculated for the NGC\,1931 region. 

\section{Conclusions} \label{conclusions}

In this study we report photometric and  polarimetric observations towards the direction of the NGC\,1931 cluster region.  
The aim of this study was to investigate the properties of dust grains in the ISM towards the direction of the clusters 
as well as the properties of intracluster dust and star formation scenario in the region. 
The following are the main conclusions of this study: 

\begin{itemize}

\item We have shown that the polarization measurements in combination with the $(U-B)-(B-V)$ 
color-color diagram provide a better identification of the cluster members. 

\item The distance to the cluster is estimated using the probable members identified on the basis of 
polarimetric and photometric data. The estimated distance of 2.3$\pm$0.3 kpc is in agreement with the 
values obtained by us in previous studies. The interstellar extinction in the cluster region is found to 
be variable with $E(B-V)_{min}~\simeq~0.5$ mag, $E(B-V)_{max}~\simeq~0.9$ mag. 
Both, the polarimetric as well as photometric studies 
indicate that the ratio of total-to-selective extinction in the cluster region, $R_{cluster}$, 
could be higher than the average.

\item The estimated mean values of $P_{V}$ and $\theta_{V}$ for the cluster NGC\,1931 are found to be 
similar to those of Stock\,8 which is located approximately at similar distance of the cluster NGC\,1931. 

\item The weighted mean of the $P_{max}$ and $\lambda_{max}$ values for NGC\,1931 are found to be 
2.51$\pm$0.03 per cent and 0.57$\pm$0.01 $\mu$m, respectively. The estimated  $\lambda_{max}$ is 
slightly higher than that for the diffuse ISM which indicates that the average size of the dust grains 
within the cluster NGC\,1931 may be slightly larger than that of the diffuse ISM. 

\item The polarization efficiency of the dust grains towards NGC\,1931 
is found to be less than that of the diffuse ISM. This could be due to the
average extinction law in the foreground medium and a differing reddening law in the intra-cluster medium.

\item The stellar density distribution in the region reveals two separate clustering in the region. 
The reddening in the northern regions is found to be variable. 
The radial extent of the region is estimated to be $\sim$3.5 arcmin.

\item The statistically cleaned CMD indicates a PMS population in the region having ages $\sim$~1$-$5~Myr. 

\item The slope of the mass function (-0.98$\pm$0.22) in the southern region  in the mass range 
0.8 $\textless$ $M/M_{\sun}$ $\textless$ 9.8  is found to be shallower in comparison to that in the 
northern region (-1.26$\pm$0.23), which is comparable to the Salpeter value (-1.35). 
The KLF of  the region is found to be comparable to the average value of slope 
\citep[$\sim$ 0.4; cf.][]{2003ARA&A..41...57L} for young clusters. However, 
the slope of the KLF is steeper in the northern region which indicates 
relatively fainter stars in the northern region as compared to the southern region.

\item The region is probably ionized by two B2 MS type stars (\#7 and \#11). 
However, the $N_{Lyc}$ derived from 1.4 GHz radio observations are quite high relative to the available 
stellar content estimated from the broad-band photometry. The maximum MS age of B2 stars could 
be $\sim$25 Myr. The YSOs identified on the basis of NIR/MIR data have masses $\sim$1-3.5$M_{\odot}$. The
mean age of the YSOs in the region is found to be 2$\pm$1 Myr which suggests
that the YSOs must be younger than the ionizing sources of the region.
The morphology of the region and ages of the YSOs and ionization sources
indicate triggered star formation in the region. 

\end{itemize}

\section*{Acknowledgments}
The authors are thankful to the anonymous referee for useful comments that improved the
content of the paper. Part of this work carried out by AKP and CE during their visits to 
National Central University, Taiwan. 
AKP and CE are thankful to the GITA/DST (India) and NSC (Taiwan) for the financial support to carry out this work.
CE is highly thankful to Dr G. Maheswar for the fruitful discussions related to polarimetry. 
This publication makes use of data products from the Two Micron All Sky Survey, 
which is a joint project of the University of Massachusetts and the Infrared Processing and 
Analysis Center/California Institute of Technology, funded by the National Aeronautics and 
Space Administration and the National Science Foundation.
This publication also makes use of data products from the Wide-field Infrared Survey Explorer, 
which is a joint project of the University of California, Los Angeles, 
and the Jet Propulsion Laboratory/California Institute of Technology, 
funded by the National Aeronautics and Space Administration.
This paper also used data from the NRAO VLA Archive Survey (NVAS). The NVAS can 
be accessed through http://www.aoc.nrao.edu/$\sim$vlbacald/. We
thank Annie Robin for letting us to use her model of stellar population synthesis.

\label{lastpage}
\bibliographystyle{apj}
\bibliography{references}


\clearpage


\begin{table}
\tiny
\caption{Observed polarzation, polarization angles and estimated Serkowski parameters.}
\label{VRI_poldata}
\begin{tabular}{lllllllllllll}\hline \hline
Star ID  & R.A  (J2000)   & DEC (J2000)  &  $P_{V} \pm \epsilon$ &  $\theta_{V} \pm \epsilon$ & $P_{Rc} \pm \epsilon$ & $\theta_{Rc} \pm \epsilon$ & $P_{Ic} \pm \epsilon $ & $\theta_{Ic} \pm \epsilon$ & $P_{max}\pm\epsilon$ & $\lambda_{max}\pm\epsilon$ &  $\sigma_{1}$ &  $\overline\epsilon$ \\
   &   (h~m~s)  & ($\degr~\arcmin~\arcsec$) &  (per cent) & ($\degr$) & (per cent) & ($\degr$) & (per cent) & ($\degr$) & (percent) & ($\mu$m)  &   &      \\
 (1)   &  (2)     &   (3)    &   (4)  &  (5)   &   (6)   &  (7)      &  (8)        &   (9)  &  (10)  &  (11) &  (12) & (13)  \\
\hline
\multicolumn{13}{c}{{\bf Stars with $V(RI)_{c}$ passband data}}\\
   3 & 5 31 26.304  & 34 11  9.838   &  2.11 $\pm$  0.09 & 153.1 $\pm$   1.1 & 2.25 $\pm$  0.08 & 151.7 $\pm$   0.9 & 1.83 $\pm$  0.09  &  150.3 $\pm$   1.4 &  2.19 $\pm$  0.06 &  0.58 $\pm$  0.04 &  1.98 &  1.22 \\
   5 & 5 31 40.241  & 34 14 26.232   &  2.29 $\pm$  0.10 & 156.3 $\pm$   1.1 & 2.39 $\pm$  0.08 & 156.1 $\pm$   1.0 & 2.16 $\pm$  0.09  &  156.2 $\pm$   1.1 &  2.37 $\pm$  0.06 &  0.61 $\pm$  0.04 &  0.61 &  0.10 \\
   8 & 5 31  9.989  & 34 11 13.049   &  0.38 $\pm$  0.11 &   6.1 $\pm$   6.4 & 0.41 $\pm$  0.09 &   2.7 $\pm$   5.2 & 0.33 $\pm$  0.11  &    0.3 $\pm$   7.3 &  0.40 $\pm$  0.08 &  0.57 $\pm$  0.25 &  0.34 &  0.50 \\
   9 & 5 31 21.278  & 34 11 17.927   &  2.82 $\pm$  0.12 & 155.6 $\pm$   1.2 & 2.71 $\pm$  0.11 & 155.8 $\pm$   1.2 & 2.57 $\pm$  0.14  &  156.7 $\pm$   1.4 &  2.82 $\pm$  0.09 &  0.58 $\pm$  0.04 &  0.67 &  0.37 \\
  10 & 5 31 43.544  & 34 16 30.803   &  0.65 $\pm$  0.16 & 172.0 $\pm$   5.8 & 0.35 $\pm$  0.14 & 163.0 $\pm$   8.8 & 0.66 $\pm$  0.13  &    4.7 $\pm$   5.0 &  --               &  --               &  --   & --    \\
  12 & 5 31 20.102  & 34 13 40.692   &  2.91 $\pm$  0.16 & 154.3 $\pm$   1.5 & 2.78 $\pm$  0.14 & 151.7 $\pm$   1.4 & 2.38 $\pm$  0.17  &  156.1 $\pm$   1.9 &  2.93 $\pm$  0.15 &  0.53 $\pm$  0.05 &  0.32 &  0.91 \\
  13 & 5 31 16.265  & 34 11 51.515   &  1.95 $\pm$  0.20 & 151.9 $\pm$   2.9 & 1.84 $\pm$  0.18 & 151.6 $\pm$   2.6 & 2.18 $\pm$  0.21  &  159.7 $\pm$   2.6 &  2.05 $\pm$  0.13 &  0.73 $\pm$  0.12 &  1.43 &  1.00 \\
  15 & 5 31 27.834  & 34 18 13.799   &  2.19 $\pm$  0.25 & 160.5 $\pm$   3.0 & 2.57 $\pm$  0.22 & 148.3 $\pm$   2.4 & 1.79 $\pm$  0.24  &  141.4 $\pm$   3.7 &  2.36 $\pm$  0.20 &  0.56 $\pm$  0.09 &  1.81 &  3.43 \\
  16 & 5 31  0.899  & 34 11 39.829   &  2.07 $\pm$  0.24 & 158.2 $\pm$   3.2 & 2.30 $\pm$  0.17 & 152.3 $\pm$   2.0 & 2.07 $\pm$  0.16  &  152.6 $\pm$   2.1 &  2.23 $\pm$  0.12 &  0.64 $\pm$  0.09 &  0.60 &  1.58 \\
  20 & 5 31  6.557  & 34  8 44.956   &  2.29 $\pm$  0.25 & 151.8 $\pm$   3.1 & 2.18 $\pm$  0.17 & 151.1 $\pm$   2.2 & 1.85 $\pm$  0.17  &  149.6 $\pm$   2.5 &  2.31 $\pm$  0.24 &  0.52 $\pm$  0.08 &  0.20 &  0.38 \\
  21 & 5 31 11.433  & 34 12 42.754   &  1.85 $\pm$  0.25 & 154.6 $\pm$   3.8 & 2.08 $\pm$  0.23 & 154.7 $\pm$   3.0 & 1.30 $\pm$  0.28  &  153.9 $\pm$   5.7 &  1.98 $\pm$  0.27 &  0.51 $\pm$  0.11 &  1.59 &  0.06 \\
  22 & 5 31 25.078  & 34 11 36.672   &  1.52 $\pm$  0.24 & 166.6 $\pm$   4.3 & 2.20 $\pm$  0.21 & 161.1 $\pm$   2.6 & 1.12 $\pm$  0.24  &  160.5 $\pm$   5.6 &  1.79 $\pm$  0.19 &  0.56 $\pm$  0.12 &  3.13 &  0.94 \\
  23 & 5 31 44.304  & 34 12  2.462   &  1.80 $\pm$  0.30 & 159.5 $\pm$   4.4 & 2.03 $\pm$  0.21 & 164.7 $\pm$   2.8 & 1.82 $\pm$  0.18  &  168.5 $\pm$   2.7 &  1.96 $\pm$  0.15 &  0.64 $\pm$  0.12 &  0.46 &  1.45 \\
  26 & 5 31 24.496  & 34  9 52.686   &  2.32 $\pm$  0.29 & 163.7 $\pm$   3.4 & 1.89 $\pm$  0.21 & 156.4 $\pm$   3.0 & 2.25 $\pm$  0.23  &  168.1 $\pm$   2.7 &  2.18 $\pm$  0.15 &  0.64 $\pm$  0.12 &  1.81 &  1.27 \\
  30 & 5 31 40.680  & 34 13 48.137   &  1.80 $\pm$  0.36 & 159.8 $\pm$   5.3 & 2.03 $\pm$  0.30 & 165.3 $\pm$   4.0 & 1.90 $\pm$  0.30  &  163.3 $\pm$   4.4 &  1.98 $\pm$  0.19 &  0.69 $\pm$  0.18 &  0.20 &  0.67 \\
  32 & 5 31 18.638  & 34 11 17.844   &  2.64 $\pm$  0.30 & 155.5 $\pm$   3.1 & 1.86 $\pm$  0.24 & 148.4 $\pm$   3.5 & 1.82 $\pm$  0.26  &  154.0 $\pm$   3.8 &  2.70 $\pm$  0.56 &  0.42 $\pm$  0.08 &  1.18 &  0.83 \\
  37 & 5 30 54.507  & 34 10 49.141   &  3.50 $\pm$  0.33 & 150.9 $\pm$   2.6 & 2.90 $\pm$  0.23 & 148.6 $\pm$   2.2 & 2.58 $\pm$  0.21  &  152.6 $\pm$   2.3 &  3.47 $\pm$  0.41 &  0.47 $\pm$  0.06 &  0.61 &  0.56 \\
  41 & 5 31 45.626  & 34 13 31.739   &  3.32 $\pm$  0.44 & 166.4 $\pm$   3.5 & 2.62 $\pm$  0.25 & 163.4 $\pm$   2.6 & 2.68 $\pm$  0.17  &  164.0 $\pm$   1.8 &  2.99 $\pm$  0.31 &  0.56 $\pm$  0.09 &  1.41 &  0.67 \\
  45 & 5 31  7.708  & 34 14 16.768   &  2.53 $\pm$  0.40 & 153.5 $\pm$   4.4 & 2.13 $\pm$  0.33 & 146.2 $\pm$   4.2 & 1.45 $\pm$  0.37  &  153.9 $\pm$   6.8 &  2.94 $\pm$  1.00 &  0.38 $\pm$  0.11 &  0.43 &  0.50 \\
  46 & 5 31 43.463  & 34 15 59.080   &  2.10 $\pm$  0.44 & 149.6 $\pm$   5.7 & 2.44 $\pm$  0.31 & 158.9 $\pm$   3.4 & 1.94 $\pm$  0.27  &  151.8 $\pm$   3.9 &  2.30 $\pm$  0.30 &  0.58 $\pm$  0.14 &  0.84 &  0.88 \\
  47 & 5 30 53.075  & 34 11 45.953   &  2.28 $\pm$  0.40 & 149.2 $\pm$   4.8 & 2.49 $\pm$  0.34 & 155.9 $\pm$   3.8 & 2.63 $\pm$  0.40  &  159.1 $\pm$   4.2 &  2.59 $\pm$  0.29 &  0.76 $\pm$  0.19 &  0.23 &  1.30 \\
  48 & 5 31 46.759  & 34 14 11.717   &  2.44 $\pm$  0.45 & 145.7 $\pm$   5.1 & 2.19 $\pm$  0.30 & 160.6 $\pm$   3.8 & 1.57 $\pm$  0.25  &  148.1 $\pm$   4.3 &  2.73 $\pm$  0.80 &  0.41 $\pm$  0.10 &  0.52 &  1.31 \\
  51 & 5 31 20.036  & 34 10  7.406   &  2.16 $\pm$  0.41 & 149.5 $\pm$   5.1 & 1.77 $\pm$  0.35 & 152.0 $\pm$   5.4 & 1.49 $\pm$  0.42  &  165.8 $\pm$   7.6 &  2.25 $\pm$  0.71 &  0.43 $\pm$  0.15 &  0.14 &  1.04 \\
  52 & 5 30 59.940  & 34 12 44.003   &  1.17 $\pm$  0.44 & 158.3 $\pm$  10.1 & 1.52 $\pm$  0.35 & 157.7 $\pm$   6.3 & 2.44 $\pm$  0.39  &  163.9 $\pm$   4.4 &    --            &   --              &       &        \\
  53 & 5 30 59.192  & 34  8 31.110   &  2.34 $\pm$  0.43 & 152.3 $\pm$   5.0 & 1.72 $\pm$  0.35 & 143.3 $\pm$   5.5 & 1.94 $\pm$  0.40  &  153.8 $\pm$   5.5 &  2.19 $\pm$  0.47 &  0.50 $\pm$  0.16 &  1.04 &  0.65 \\
  55 & 5 30 59.360  & 34 14 20.011   &  2.25 $\pm$  0.47 & 155.5 $\pm$   5.7 & 2.45 $\pm$  0.41 & 158.8 $\pm$   4.7 & 2.21 $\pm$  0.49  &  158.3 $\pm$   6.1 &  2.39 $\pm$  0.27 &  0.64 $\pm$  0.20 &  0.21 &  0.37 \\
  57 & 5 31 20.978  & 34 13 47.093   &  3.18 $\pm$  0.43 & 155.7 $\pm$   3.8 & 2.47 $\pm$  0.34 & 152.9 $\pm$   3.8 & 2.51 $\pm$  0.37  &  141.6 $\pm$   4.1 &  3.04 $\pm$  0.51 &  0.49 $\pm$  0.11 &  0.97 &  1.44 \\
  58 & 5 30 53.382  & 34 12 17.334   &  2.67 $\pm$  0.47 & 155.0 $\pm$   4.9 & 2.52 $\pm$  0.33 & 151.7 $\pm$   3.6 & 2.47 $\pm$  0.31  &  149.1 $\pm$   3.4 &  2.64 $\pm$  0.27 &  0.61 $\pm$  0.14 &  0.37 &  0.77 \\
  71 & 5 31  1.710  & 34  8 26.826   &  2.88 $\pm$  0.60 & 156.6 $\pm$   5.8 & 2.13 $\pm$  0.47 & 155.1 $\pm$   6.0 & 2.60 $\pm$  0.56  &   -3.3 $\pm$   5.9 &  2.63 $\pm$  0.48 &  0.56 $\pm$  0.20 &  1.12 &  9.12 \\
  73 & 5 31 37.341  & 34 16 48.904   &  2.28 $\pm$  0.64 & 153.5 $\pm$   7.6 & 2.49 $\pm$  0.52 & 159.6 $\pm$   5.7 & 2.32 $\pm$  0.50  &  162.3 $\pm$   5.9 &  2.45 $\pm$  0.33 &  0.66 $\pm$  0.24 &  0.11 &  0.78 \\
  79 & 5 31  3.560  & 34 15 20.297   &  4.76 $\pm$  0.55 & 151.8 $\pm$   3.3 & 5.05 $\pm$  0.47 & 135.2 $\pm$   2.6 & 4.81 $\pm$  0.53  &  167.0 $\pm$   3.1 &  5.04 $\pm$  0.31 &  0.66 $\pm$  0.11 &  0.05 &  3.53 \\
  84 & 5 31 17.663  & 34 13 27.548   &  2.30 $\pm$  0.57 & 157.1 $\pm$   6.8 & 2.89 $\pm$  0.48 & 158.1 $\pm$   4.6 & 1.79 $\pm$  0.57  &  148.9 $\pm$   8.7 &  2.56 $\pm$  0.47 &  0.55 $\pm$  0.20 &  1.23 &  0.46 \\
  92 & 5 31 20.749  & 34 11 53.851   &  1.98 $\pm$  0.61 & 167.9 $\pm$   8.5 & 2.45 $\pm$  0.47 & 156.0 $\pm$   5.2 & 1.55 $\pm$  0.49  &  136.9 $\pm$   8.6 &  2.23 $\pm$  0.58 &  0.53 $\pm$  0.21 &  1.04 &  1.92 \\
  93 & 5 31 37.064  & 34 14 15.670   &  2.28 $\pm$  0.69 & 146.6 $\pm$   8.4 & 4.05 $\pm$  0.58 & 158.8 $\pm$   4.0 & 3.02 $\pm$  0.56  &  160.9 $\pm$   5.1 &  3.40 $\pm$  0.50 &  0.79 $\pm$  0.25 &  1.65 &  1.51 \\
 103 & 5 31 13.736  & 34 13 26.782   &  3.14 $\pm$  0.70 & 154.0 $\pm$   6.2 & 3.36 $\pm$  0.57 & 147.1 $\pm$   4.7 & 2.08 $\pm$  0.66  &  142.9 $\pm$   8.5 &  3.43 $\pm$  0.97 &  0.47 $\pm$  0.16 &  0.97 &  0.93 \\ 
\multicolumn{13}{c}{{\bf Stars embedded in the nebulosity}}\\
   7 & 5 31 27.071  &  34 14 49.520  &  0.82 $\pm$  0.11 & 148.9 $\pm$   3.3 & 0.73 $\pm$  0.09 & 131.8 $\pm$   3.1 & 0.55 $\pm$  0.09 &    99.4 $\pm$   3.9  &  0.88 $\pm$  0.18 &  0.43 $\pm$  0.09 &  0.36 &  6.47 \\
  11 & 5 31 26.523  &  34 14 45.071  &  0.62 $\pm$  0.13 & 162.9 $\pm$   5.0 & 0.86 $\pm$  0.10 & 170.6 $\pm$   2.9 & 0.87 $\pm$  0.10 &    26.7 $\pm$   2.8  &  0.90 $\pm$  0.13 &  0.89 $\pm$  0.20 &  0.56 & 13.42 \\
  14 & 5 31 26.434  &  34 14 56.224  &  2.11 $\pm$  0.22 & 140.4 $\pm$   2.9 & 2.10 $\pm$  0.12 & 143.6 $\pm$   1.5 & 1.30 $\pm$  0.19 &   127.8 $\pm$   3.9  &  2.40 $\pm$  0.36 &  0.44 $\pm$  0.07 &  2.18 &  1.89 \\
  34 & 5 31 27.504  &  34 15  3.132  &  2.11 $\pm$  0.37 & 156.0 $\pm$   4.6 & 1.11 $\pm$  0.29 & 135.0 $\pm$   7.0 & 1.04 $\pm$  0.29 &   175.0 $\pm$   7.4  &  --  &   --  &  --  &  -- \\
  35 & 5 31 25.024  &  34 14 40.484  &  6.51 $\pm$  0.32 & 159.1 $\pm$   1.3 & 6.23 $\pm$  0.23 & 163.3 $\pm$   1.1 & 6.82 $\pm$  0.20 &   163.4 $\pm$   0.8  &  6.73 $\pm$  0.14 &  0.71 $\pm$  0.04 &  2.59 &  2.64 \\
  49 & 5 31 25.012  &  34 14 17.387  &  1.61 $\pm$  0.49 & 158.2 $\pm$   8.0 & 1.58 $\pm$  0.40 & 135.6 $\pm$   7.0 & 1.73 $\pm$  0.41 &   130.0 $\pm$   6.4  &  1.69 $\pm$  0.27 &  0.71 $\pm$  0.30 &  0.35 &  2.39 \\
  68 & 5 31 31.881  &  34 13 57.500  &  2.03 $\pm$  0.57 & 152.1 $\pm$   7.6 & 1.70 $\pm$  0.42 & 149.6 $\pm$   6.8 & 1.52 $\pm$  0.41 &   156.9 $\pm$   7.3  &  2.01 $\pm$  0.71 &  0.48 $\pm$  0.20 &  0.20 &  0.34 \\
  70 & 5 31 26.662  &  34 14 21.656  &  2.21 $\pm$  0.60 & 142.5 $\pm$   7.6 & 3.25 $\pm$  0.49 & 145.9 $\pm$   4.2 & 2.32 $\pm$  0.50 &   141.4 $\pm$   5.9  &  2.74 $\pm$  0.31 &  0.67 $\pm$  0.21 &  1.37 &  0.25 \\
  72 & 5 31 20.975  &  34 14 22.715  &  2.08 $\pm$  0.60 & 163.3 $\pm$   7.5 & 2.67 $\pm$  0.40 & 152.8 $\pm$   4.1 & 2.82 $\pm$  0.33 &   142.4 $\pm$   3.2  &  2.86 $\pm$  0.45 &  0.88 $\pm$  0.26 &  0.14 &  2.12 \\
  88 & 5 31 25.693  &  34 14 41.104  &  8.66 $\pm$  0.47 & 164.7 $\pm$   1.5 & 6.36 $\pm$  0.35 & 153.0 $\pm$   1.6 & 6.42 $\pm$  0.33 &   147.8 $\pm$   1.4  &  8.26 $\pm$  0.61 &  0.47 $\pm$  0.04 &  2.88 &  6.41 \\
  99 & 5 31 25.777  &  34 14 33.565  &  4.77 $\pm$  0.64 & 164.3 $\pm$   3.6 & 4.80 $\pm$  0.52 & 154.7 $\pm$   3.0 & 5.39 $\pm$  0.53 &   162.9 $\pm$   2.7  &  5.22 $\pm$  0.40 &  0.76 $\pm$  0.14 &  0.84 &  1.17 \\
 139 & 5 31 20.503  &  34 15 17.975  &  4.51 $\pm$  1.01 & 127.6 $\pm$   6.8 & 3.18 $\pm$  0.60 & 146.4 $\pm$   5.3 & 4.26 $\pm$  0.44 &   143.3 $\pm$   2.8  &  4.05 $\pm$  0.37 &  0.77 $\pm$  0.22 &  1.74 &  2.32 \\
\multicolumn{13}{c}{{\bf Stars with either single or double band data}}\\
   1 & 5 30 59.954  & 34 15 32.065  & 1.97 $\pm$  0.08 & 150.7 $\pm$   1.1  & --                & --                &  1.61 $\pm$  0.05 & 151.5 $\pm$   0.9  &  -- & -- & -- & --  \\
  60 & 5 31 44.919  & 34 12 17.676  & 2.16 $\pm$  0.52 & 156.5 $\pm$   6.4  &  2.11 $\pm$  0.40 & 164.9 $\pm$   5.2 &     --            & --                 &  -- & -- & -- & --  \\ 
  65 & 5 31 25.097  & 34 10 24.942  & 2.70 $\pm$  0.48 & 161.1 $\pm$   4.9  &  1.88 $\pm$  0.41 & 163.9 $\pm$   5.9 &     --            & --                 &  -- & -- & -- & --  \\
  66 & 5 31 12.905  & 34 13 54.822  & 2.06 $\pm$  0.54 & 162.4 $\pm$   7.2  &  1.83 $\pm$  0.47 & 154.5 $\pm$   7.0 &     --            & --                 &  -- & -- & -- & --  \\
  76 & 5 31 23.673  & 34 10 40.710  & 3.04 $\pm$  0.56 & 159.7 $\pm$   5.1  &  2.06 $\pm$  0.49 & 154.3 $\pm$   6.5 &     --            & --                 &  -- & -- & -- & --  \\
  85 & 5 31 27.167  & 34 18 19.141  &   --             &  --                &       --          &      --           &  2.99 $\pm$  0.28 & 143.9 $\pm$   2.5  &  -- & -- & -- & --  \\
  86 & 5 31 36.690  & 34 16 58.282  &   --             &  --                &       --          &      --           &  1.51 $\pm$  0.45 & 167.0 $\pm$   7.9  &  -- & -- & -- & --  \\
 102 & 5 31 35.799  & 34 13 22.894  &   --             &  --                &  2.87 $\pm$  0.59 & 179.1 $\pm$   5.7 &  2.79 $\pm$  0.59 & 154.8 $\pm$   5.8  &  -- & -- & -- & --  \\
 115 & 5 31 26.544  & 34 16  0.836  & 3.11 $\pm$  0.90 & 168.7 $\pm$   7.7  &  2.60 $\pm$  0.45 & 171.3 $\pm$   4.7 &      --           &      --            &  -- & -- & -- & --  \\
 128 & 5 31 23.382  & 34 12 11.347  &   --             &         --         &  2.23 $\pm$  0.72 & 154.4 $\pm$   8.9 &      --           &      --            &  -- & -- & -- & --  \\
 133 & 5 31 30.090  & 34 16 47.089  & 3.30 $\pm$  0.95 & 147.1 $\pm$   8.1  &     --            &     --            &      --           &      --            &  -- & -- & -- & --  \\
 236 & 5 31 23.741  & 34 13 47.338  & 6.52 $\pm$  0.55 & 130.1 $\pm$   2.5  &  3.59 $\pm$  0.41 & 129.4 $\pm$   3.2 &      --           &      --            &  -- & -- & -- & --  \\
2001$^\dagger$ & 5 30 47.429  & 34 11 47.864  & --              &       --           &  1.96 $\pm$  0.61 & 158.3 $\pm$   8.4 &  2.58 $\pm$  0.70 & 169.6 $\pm$   7.5   &  -- & -- & -- & --  \\
2002$^\dagger$ & 5 30 49.688  & 34 11 28.666  & 3.10 $\pm$  0.59 & 147.1 $\pm$   5.3  &  3.54 $\pm$  0.51 & 147.8 $\pm$   4.1 &      --           &       --           &  -- & -- & -- & --  \\
2003$^\dagger$ & 5 31 24.224  & 34 14  0.139  & 3.16 $\pm$  0.79 & 172.6 $\pm$   6.6  &      --           &          --       &      --           &        --          &  -- & -- & -- & --  \\  
\hline \hline
\end{tabular}\\
$^\dagger$: Optical data is not available for the three stars, namely \#2001, \#2002 and \#2003. Their ids are 
given arbitrarily. 
\end{table}


\begin{table}
\centering
\tiny
\caption{The optical (present work), $JHK_{s}$ (2MASS) data for the identified YSOs}
\label{whole_data_ysos1}
\begin{tabular}{ccccccccccc}\hline \hline
ID       &        RA (J2000)  &    Dec (J2000)  &  U$\pm\sigma_{U}$  &   B$\pm\sigma_{B}$  &   V$\pm\sigma_{V}$  &   $R_{c}$$\pm\sigma_{R_{c}}$  &   $I_{c}$$\pm\sigma_{I_{c}}$  &   $J\pm\sigma_{J}$  &  $H\pm\sigma_{H}$  &  $K\pm\sigma_{K}$   \\ 
   &    (h~m~s)      &    ($\degr~\arcmin~\arcsec$)    &    (mag)   &   (mag)   &  (mag)   &  (mag)   &  (mag)  &  (mag)   & (mag)  & (mag)  \\

(1) & (2) & (3) & (4) & (5) & (6) & (7) & (8) & (9) & (10) & (11)   \\
\hline
\hline 
    46  &  5 31 43.461   & 34 15 59.011  &   16.689  $\pm$    0.007  &   15.684  $\pm$    0.004  &   14.436  $\pm$    0.002  &   13.719  $\pm$    0.002  &   13.071  $\pm$    0.003  &   12.087  $\pm$    0.022  &   11.500  $\pm$    0.021  &   11.331  $\pm$    0.019  \\ 
    58  &  5 30 53.386   & 34 12 17.417  &   16.964  $\pm$    0.014  &   16.098  $\pm$    0.004  &   14.784  $\pm$    0.004  &   13.998  $\pm$    0.003  &   13.241  $\pm$    0.003  &   12.060  $\pm$    0.019  &   11.433  $\pm$    0.022  &   11.237  $\pm$    0.018  \\ 
    72  &  5 31 20.999   & 34 14 23.129  &   16.685  $\pm$    0.008  &   16.223  $\pm$    0.004  &   15.027  $\pm$    0.003  &   14.237  $\pm$    0.003  &   13.432  $\pm$    0.003  &   12.119  $\pm$    0.023  &   11.449  $\pm$    0.023  &   11.220  $\pm$    0.019  \\ 
    85  &  5 31 27.171   & 34 18 19.087  &   18.808  $\pm$    0.020  &   17.055  $\pm$    0.009  &   15.183  $\pm$    0.010  &   14.016  $\pm$    0.004  &   12.940  $\pm$    0.009  &   11.192  $\pm$    0.022  &   10.203  $\pm$    0.021  &    9.897  $\pm$    0.018  \\ 
    88  &  5 31 36.688   & 34 16 57.950  &   15.893  $\pm$    0.093  &   15.916  $\pm$    0.090  &   15.274  $\pm$    0.026  &   14.662  $\pm$    0.060  &   13.748  $\pm$    0.150  &   13.406  $\pm$    0.037  &   12.924  $\pm$    0.035  &   12.818  $\pm$    0.033  \\ 
   129  &  5 31 32.081   & 34  9 26.896  &   17.474  $\pm$    0.009  &   16.924  $\pm$    0.018  &   15.965  $\pm$    0.004  &   15.268  $\pm$    0.011  &   14.551  $\pm$    0.009  &   13.252  $\pm$    0.022  &   12.688  $\pm$    0.025  &   12.172  $\pm$    0.021  \\ 
   139  &  5 31 20.503   & 34 15 17.975  &   18.715  $\pm$    0.018  &   17.715  $\pm$    0.011  &   16.094  $\pm$    0.005  &   15.016  $\pm$    0.012  &   13.943  $\pm$    0.005  &   12.120  $\pm$    0.022  &   11.221  $\pm$    0.021  &   10.919  $\pm$    0.018  \\ 
   236  &  5 31 23.731   & 34 13 47.352  &   18.290  $\pm$    0.026  &   18.168  $\pm$    0.022  &   16.980  $\pm$    0.029  &   16.140  $\pm$    0.015  &   15.032  $\pm$    0.236  &   12.805  $\pm$    0.023  &   12.057  $\pm$    0.024  &   11.269  $\pm$    0.018  \\ 
   544  &  5 31 19.522   & 34 14 41.392  &                    --                   &   20.354  $\pm$    0.018  &   18.623  $\pm$    0.022  &   17.524  $\pm$    0.015  &   16.420  $\pm$    0.111  &   14.665  $\pm$    0.034  &   13.699  $\pm$    0.033  &   13.267  $\pm$    0.029  \\ 
   589  &  5 31 23.966   & 34 12 21.672  &                    --                   &   20.661  $\pm$    0.021  &   18.776  $\pm$    0.011  &   17.694  $\pm$    0.012  &   16.618  $\pm$    0.048  &   14.877  $\pm$    0.040  &   13.947  $\pm$    0.046  &   13.636  $\pm$    0.044  \\ 
   609  &  5 31 24.859   & 34 13 34.457  &                    --                   &   20.444  $\pm$    0.044  &   18.876  $\pm$    0.012  &   17.583  $\pm$    0.021  &   16.401  $\pm$    0.070  &   14.489  $\pm$    0.034  &   13.549  $\pm$    0.037  &   13.135  $\pm$    0.034  \\ 
  1493  &  5 31  4.709   & 34 17  7.112  &                    --                   &                    --                   &   21.161  $\pm$    0.048  &   19.845  $\pm$    0.029  &   18.253  $\pm$    0.021  &   16.274  $\pm$    0.096  &   15.404  $\pm$    0.102  &   14.799  $\pm$    0.089  \\ 
     a  &  5 31 23.926   & 34  9 56.668  &                    --                   &                    --                   &                    --                   &                    --                   &                    --                   &   14.737  $\pm$    0.041  &   13.021  $\pm$    0.049  &   11.750  $\pm$    0.021  \\ 
     b  &  5 30 50.131   & 34 12 57.532  &                    --                   &                    --                   &                    --                   &                    --                   &                    --                   &   13.392  $\pm$    0.017  &   12.800  $\pm$    0.022  &   12.541  $\pm$    0.021  \\ 
     c  &  5 31 38.609   & 34 14 24.760  &                    --                   &                    --                   &                    --                   &                    --                   &                    --                   &   15.171  $\pm$    0.042  &   14.112  $\pm$    0.035  &   13.482  $\pm$    0.033  \\ 
     d  &  5 30 47.194   & 34  7  0.048  &                    --                   &                    --                   &                    --                   &                    --                   &                    --                   &   16.188  $\pm$    0.106  &   15.204  $\pm$    0.096  &   14.597  $\pm$    0.091  \\
\hline
\hline
\end{tabular}\\
Note: The stars (a, b, c, d) are not covered by optical photometric observations.\\
\end{table}

\begin{table}
\centering
\tiny
\caption{The 3.6 $\mu$m and 4.5 $\mu$m ({\it SPITZER}), 3.4 $\mu$m and 4.6 $\mu$m, 12 $\mu$m and 22 $\mu$m (WISE) data for the identified YSOs}
\label{whole_data_ysos2}
\begin{tabular}{ccccccccc}\hline \hline
ID       &        RA (J2000)  &    Dec (J2000)  &  $3.6\pm\sigma_{3.6}\mu$m  & $4.5\pm\sigma_{4.5}\mu$m  &  $3.35\pm\sigma_{3.35}\mu$m  & $4.6\pm\sigma_{4.6}\mu$m &  $11.6\pm\sigma_{11.6}\mu$m &   $22\pm\sigma_{22}\mu$m \\ 
   &    (h~m~s)      &    ($\degr~\arcmin~\arcsec$)    &    (mag)   &   (mag)   &  (mag)   &  (mag)   &  (mag)  &  (mag)   \\
(1) & (2) & (3) & (4) & (5) & (6) & (7) & (8) & (9)   \\
\hline
\hline
    46  &  5 31 43.461  & 34 15 59.011  &                  --                    &                  --                    &   11.239  $\pm$    0.026  &   11.283  $\pm$    0.026  &    8.709  $\pm$    0.047  &    6.661  $\pm$    0.133  \\ 
    58  &  5 30 53.386  & 34 12 17.417  &                  --                    &                  --                    &   11.125  $\pm$    0.023  &   11.170  $\pm$    0.022  &   11.391       &    8.956     \\ 
    72  &  5 31 20.999  & 34 14 23.129  &                  --                    &                  --                    &   10.824  $\pm$    0.024  &   10.676  $\pm$    0.020  &    7.212  $\pm$    0.020  &    2.755  $\pm$    0.021  \\ 
    85  &  5 31 27.171  & 34 18 19.087  &                  --                    &                  --                    &    9.657  $\pm$    0.024  &    9.707  $\pm$    0.021  &    9.634       &    6.990       \\ 
    88  &  5 31 36.688  & 34 16 57.950  &                  --                    &                  --                    &   12.414  $\pm$    0.026  &   12.337  $\pm$    0.029  &    8.587  $\pm$    0.059  &    6.010  $\pm$    0.079  \\ 
   129  &  5 31 32.081  & 34  9 26.896  &   10.680  $\pm$    0.029  &   10.035  $\pm$    0.027  &   10.802  $\pm$    0.030  &    9.914  $\pm$    0.025  &    6.321  $\pm$    0.049  &    4.511  $\pm$    0.103  \\ 
   139  &  5 31 20.503  & 34 15 17.975  &   10.674  $\pm$    0.039  &   10.646  $\pm$    0.025  &   10.372  $\pm$    0.023  &   10.262  $\pm$    0.020  &    5.958  $\pm$    0.018  &    2.143  $\pm$    0.022  \\ 
   236  &  5 31 23.731  & 34 13 47.352  &   10.113  $\pm$    0.024  &    9.580  $\pm$    0.026  &   10.121  $\pm$    0.026  &    9.392  $\pm$    0.022  &    6.302  $\pm$    0.036  &    3.664  $\pm$    0.079  \\ 
   544  &  5 31 19.522  & 34 14 41.392  &   12.350  $\pm$    0.030  &   12.047  $\pm$    0.032  &   12.462  $\pm$    0.094  &   11.938  $\pm$    0.107  &    8.016  $\pm$    0.133  &    3.966  $\pm$    0.104  \\ 
   589  &  5 31 23.966  & 34 12 21.672  &   12.823  $\pm$    0.045  &   12.531  $\pm$    0.052  &   12.864  $\pm$    0.076  &   12.565  $\pm$    0.101  &    8.469  $\pm$    0.150  &    6.334  $\pm$    0.441  \\ 
   609  &  5 31 24.859  & 34 13 34.457  &   12.457  $\pm$    0.045  &   12.119  $\pm$    0.046  &   12.382  $\pm$    0.069  &   11.633  $\pm$    0.073  &    7.946  $\pm$    0.124  &    3.769  $\pm$    0.057  \\ 
  1493  &  5 31  4.709  & 34 17  7.112  &   14.147  $\pm$    0.052  &   13.604  $\pm$    0.049  &   14.192  $\pm$    0.062  &   13.540  $\pm$    0.092  &    9.274  $\pm$    0.087  &    7.129  $\pm$    0.324  \\ 
   a    &  5 31 23.926  & 34  9 56.668  &   10.085  $\pm$    0.033  &    9.489  $\pm$    0.022  &   10.215  $\pm$    0.033  &    9.341  $\pm$    0.028  &    6.734  $\pm$    0.130  &    4.476  $\pm$    0.420  \\ 
   b    &  5 30 50.131  & 34 12 57.532  &   12.045  $\pm$    0.025  &   11.539  $\pm$    0.037  &   12.119  $\pm$    0.027  &   11.536  $\pm$    0.023  &    7.976  $\pm$    0.026  &    5.753  $\pm$    0.040  \\ 
   c    &  5 31 38.609  & 34 14 24.760  &   12.858  $\pm$    0.036  &   12.428  $\pm$    0.036  &   12.889  $\pm$    0.109  &   12.350  $\pm$    0.124  &    8.218  $\pm$    0.231  &    5.451  $\pm$    0.258  \\ 
   d    &  5 30 47.194  & 34  7  0.048  &                  --       &                  --       &   13.685  $\pm$    0.031  &   13.461  $\pm$    0.042  &    9.948  $\pm$    0.079  &    7.980  $\pm$    0.196  \\ 
\hline
\hline
\end{tabular}\\
Note: The stars (a, b, c, d) are not covered by optical photometric observations.\\
\end{table}

\begin{table}
\centering
\tiny
\caption{Age and mass of YSOs obtained in the present study}
\label{age_and_mass_neelam} 
\begin{tabular}{cccccc}
\hline 
ID    &  Age $\pm \sigma$ & Mass $\pm$ $\sigma$   & ID    &  Age $\pm \sigma$ & Mass $\pm$ $\sigma$  \\
     &         (Myr)             &     ($M_{\sun}$)   &         &         (Myr)             &     ($M_{\sun}$)     \\
\hline
    32  &    2.3  $\pm$   0.3  &    3.0  $\pm$   0.1   &    1009  &   $\textgreater$ 5.0  &    0.7  $\pm$   0.1  \\ 
    46  &    1.2  $\pm$   0.3  &    3.4  $\pm$   0.2   &    1017  &   $\textgreater$ 5.0   &    0.7  $\pm$   0.1  \\ 
    58  &    0.5  $\pm$   0.1  &    3.5  $\pm$   0.1   &    1056  &    1.6  $\pm$   0.5  &    0.5  $\pm$   0.1  \\ 
    72  &    0.5  $\pm$   0.1  &    3.1  $\pm$   0.2   &    1074  &    2.4  $\pm$   1.5  &    0.5  $\pm$   0.1  \\ 
    88  &    0.9  $\pm$   0.7  &    2.8  $\pm$   0.2   &    1133  &    1.2  $\pm$   0.1  &    0.4  $\pm$   0.1  \\ 
    98  &    0.8  $\pm$   1.2  &    2.2  $\pm$   0.3   &    1188  &    1.5  $\pm$   0.2  &    0.4  $\pm$   0.1  \\ 
   129  &    2.7  $\pm$   0.9  &    2.4  $\pm$   0.2   &    1200  &    $\textgreater$ 5.0   &    0.6  $\pm$   0.1  \\ 
   236  &    0.4  $\pm$   0.3  &    1.0  $\pm$   0.3   &    1204  &    $\textgreater$ 5.0   &    0.6  $\pm$   0.1  \\ 
   298  &    2.2  $\pm$   0.9  &    1.5  $\pm$   0.1   &    1223  &    2.7  $\pm$   0.5  &    0.5  $\pm$   0.1  \\ 
   321  &    3.9  $\pm$   0.6  &    1.6  $\pm$   0.1   &    1238  &    1.5  $\pm$   0.3  &    0.4  $\pm$   0.1  \\ 
   544  &    0.9  $\pm$   0.3  &    0.7  $\pm$   0.1   &    1268  &    2.5  $\pm$   1.3  &    0.4  $\pm$   0.1  \\ 
   589  &    1.3  $\pm$   0.2  &    0.7  $\pm$   0.1   &    1279  &    1.8  $\pm$   0.2  &    0.4  $\pm$   0.1  \\ 
   609  &    0.6  $\pm$   0.1  &    0.5  $\pm$   0.1   &    1281  &    3.4  $\pm$   0.9  &    0.5  $\pm$   0.1  \\ 
   668  &    1.7  $\pm$   0.3  &    0.7  $\pm$   0.1   &    1298  &    1.4  $\pm$   0.1  &    0.3  $\pm$   0.1  \\ 
   671  &    2.3  $\pm$   0.9  &    0.8  $\pm$   0.1   &    1349  &    1.4  $\pm$   0.1  &    0.3  $\pm$   0.1  \\ 
   719  &    3.1  $\pm$   1.0  &    0.8  $\pm$   0.1   &    1387  &    1.6  $\pm$   0.1  &    0.3  $\pm$   0.1  \\ 
   737  &    3.7  $\pm$   0.8  &    0.9  $\pm$   0.1   &    1395  &    3.7  $\pm$   0.7  &    0.4  $\pm$   0.1  \\ 
   745  &    2.1  $\pm$   0.8  &    0.7  $\pm$   0.1   &    1448  &    4.8  $\pm$   0.2  &    0.5  $\pm$   0.1  \\ 
   953  &    2.8  $\pm$   1.0  &    0.6  $\pm$   0.1   &    1481  &    1.6  $\pm$   0.1  &    0.3  $\pm$   0.1 \\ 
   967  &    3.2  $\pm$   0.9  &    0.6  $\pm$   0.1   &    1493  &    1.8  $\pm$   0.2  &    0.3  $\pm$   0.1  \\ 
        &                      &                       &    1575  &    2.4  $\pm$   0.3  &    0.3  $\pm$   0.1  \\ 
\hline \hline
\end{tabular}
\end{table}

\begin{table}
\centering
\tiny
\caption{The physical parameters of the YSOs estimated from the SED fits}
\label{age_and_mass_manash}
\begin{tabular}{cccccc}
\hline\hline
 ID & \multicolumn{1}{c}{M$_{\ast}$} &\multicolumn{1}{c}{T$_{\ast}$ } & \multicolumn{1}{c}{age}  & \multicolumn{1}{c}{$\dot{M}_{\rm acc}$ }  & \multicolumn{1}{c}{A$_V$ }  \\
    & \multicolumn{1}{c}{($M_\odot$)} & \multicolumn{1}{c}{(10$^{3}$ K)} &\multicolumn{1}{c}{(10$^{6}$ yr)}  & \multicolumn{1}{c}{(10$^{-7}$ $M_\odot$/yr)}  & \multicolumn{1}{c}{(mag)}   \\
 \hline
 \hline
    46  &    3.44  $\pm$    0.20  &    5.30  $\pm$    0.45  &    0.980  $\pm$    0.352  &    0.011  $\pm$    0.000  &    1.4  $\pm$    0.4  \\ 
    58  &    3.65  $\pm$    0.17  &    5.33  $\pm$    0.05  &    0.901  $\pm$    0.099  &    0.000  $\pm$    0.000  &    1.8  $\pm$    0.1  \\ 
    72  &    3.07  $\pm$    0.44  &    6.02  $\pm$    1.05  &    1.532  $\pm$    0.588  &    1.036  $\pm$    0.028  &    2.1  $\pm$    0.6  \\ 
    85  &    3.74  $\pm$    0.00  &    5.35  $\pm$    0.00  &    0.852  $\pm$    0.000  &    0.000  $\pm$    0.000  &    2.9  $\pm$    0.0  \\ 
    88  &    3.45  $\pm$    0.49  &   12.81  $\pm$    1.82  &    4.307  $\pm$    2.346  &    0.000  $\pm$    0.000  &    2.8  $\pm$    0.4  \\ 
   129  &    4.34  $\pm$    0.66  &   14.89  $\pm$    1.91  &    3.667  $\pm$    0.514  &    0.006  $\pm$    0.001  &    4.0  $\pm$    0.4  \\ 
   139  &    3.56  $\pm$    0.51  &    5.70  $\pm$    0.57  &    1.035  $\pm$    0.408  &    0.821  $\pm$    0.032  &    3.1  $\pm$    1.0  \\ 
   236  &    3.76  $\pm$    0.43  &   13.70  $\pm$    1.10  &    5.052  $\pm$    2.050  &    0.082  $\pm$    0.007  &    4.1  $\pm$    0.4  \\ 
   544  &    2.07  $\pm$    0.71  &    6.61  $\pm$    2.80  &    4.171  $\pm$    2.903  &    0.206  $\pm$    0.006  &    3.9  $\pm$    1.4  \\ 
   589  &    2.23  $\pm$    0.27  &    7.73  $\pm$    1.83  &    5.755  $\pm$    2.119  &    0.046  $\pm$    0.002  &    4.8  $\pm$    1.3  \\ 
   609  &    1.76  $\pm$    0.76  &    5.56  $\pm$    2.20  &    2.994  $\pm$    2.648  &    0.177  $\pm$    0.006  &    3.5  $\pm$    1.3  \\ 
  1493  &    2.07  $\pm$    0.60  &    5.20  $\pm$    0.61  &    3.261  $\pm$    1.230  &    0.155  $\pm$    0.003  &    3.1  $\pm$    0.6  \\ 
a       &    3.95  $\pm$	1.25 & 13.07	$\pm$	3.67 & 4.526	$\pm$	2.497 &  0.265	 $\pm$	0.005 &  10.0  $\pm$	1.7  \\
b       &    2.47	$\pm$	0.64 & 7.27	$\pm$	2.93 & 4.330	$\pm$	1.314 &  0.024	 $\pm$	0.002 &  3.4	$\pm$	0.5  \\
c       &    2.85	$\pm$	1.30 & 9.97	$\pm$	4.24 & 4.202	$\pm$	2.264 &  0.107	 $\pm$	0.004 &  8.7	$\pm$	2.3  \\
d       &    2.02	$\pm$	0.67 & 8.09	$\pm$	2.73 & 5.991	$\pm$	2.842 &  0.070	 $\pm$	0.002 &  8.8	$\pm$	2.8  \\
\hline
\hline
\end{tabular} \\
Note: The stars (a, b, c, d) are not covered by optical photometric observations. \\
\end{table}

\end{document}